\documentclass[11pt,a4paper]{article}
\usepackage{jheppub}
\usepackage[numbers,sort&compress]{natbib}
\usepackage{xcolor}
\usepackage{amsmath}
\usepackage{amssymb}
\usepackage{amsfonts}
\usepackage{tocloft}
\usepackage{hyperref}
\usepackage{graphicx}
\usepackage{shuffle}
\newcommand{\hs}{\hypersetup{pdftitle={PolyLogTools},pdfcreator={Claude Duhr, Falko Dulat},colorlinks=true,linkcolor=[rgb]{0.15,0.35,0.75},citecolor=[rgb]{0.675,0,0.2},urlcolor=[rgb]{0.15,0.35,0.65}}}
\hs
\newcommand{\polylogtools}{{\sc PolyLogTools}}
\newcommand{\plt}{{\sc PolyLogTools}}
\newcommand{\mathematica}{{\sc Mathematica}}
\newcommand{\hpl}{{\sc Hpl}}
\newcommand{\ginac}{{\sc GiNaC}}

\newcommand{\exbox}[1]{\begin{center}\fbox{\parbox[c]{14.8cm}{#1}}\end{center}}
\newcommand{\inline}[2]{\begin{tabular}{rl}{\tt \phantom{a}In[#1]:= } #2\end{tabular}}
\newcommand{\outline}[2]{\begin{tabular}{rl}{\tt Out[#1]:= } #2\end{tabular}}
\newcommand{\breakline}{\rule{14.8cm}{0.5pt}}

\definecolor{lightgrey}{RGB}{235,245,255}
\def\mybtab#1\myetab{\begin{tabular}{p{61mm}p{79mm}}#1\end{tabular}}
\def\btab#1\etab{\begin{tabular}{p{55mm}p{85mm}}#1\end{tabular}}
\def\btabx#1\etabx{\begin{tabular}{p{60mm}p{50mm}}#1\end{tabular}}
\def\btaby#1\etaby{\begin{tabular}{p{15mm}p{95mm}}#1\end{tabular}}
\def\bcen{\begin{center}\begin{small}}
\def\ecen{\end{small}\end{center}}
\def\mybgfb#1\myegfb{\bcen\fcolorbox{black}{lightgrey}{\parbox{148mm}{\mybtab#1\myetab}}\ecen}
\def\bgfb#1\egfb{\bcen\fcolorbox{black}{lightgrey}{\parbox{148mm}{\btab#1\etab}}\ecen}
\def\bgfbx#1\egfbx{\bcen\fcolorbox{black}{lightgrey}{\parbox{118mm}{\btabx#1\etabx}}\ecen}
\def\bgfbalign#1\egfbalign{\bcen\fcolorbox{black}{lightgrey}{\parbox{118mm}{\btaby#1\etaby}}\ecen}
\def\beq{\begin{equation}}
\def\eeq{\end{equation}}
\def\bsp#1\esp{\begin{split}#1\end{split}}
\newcommand{\li}[1]{\textrm{Li}_{#1}}

\newcommand{\cS}{{\cal S}}

\newcommand{\cG}{\mathcal{G}}
\newcommand{\cC}{\mathcal{C}}

\title{PolyLogTools -- Polylogs for the masses}
\author[a,b]{Claude Duhr}
\author[c]{Falko Dulat}

\affiliation[a]{Theoretical Physics Department, CERN, Geneva, Switzerland} 
\affiliation[b]{Center for Cosmology, Particle Physics and Phenomenology (CP3),\\
Universit\'e Catholique de Louvain, 1348 Louvain-La-Neuve, Belgium}
\affiliation[c]{SLAC National Accelerator Laboratory, Stanford University, Stanford, CA 94309, USA}

\emailAdd{claude.duhr@cern.ch}
\emailAdd{dulatf@slac.stanford.edu}

\abstract{
 We review recent developments in the study of multiple polylogarithms, including the Hopf algebra of the multiple polylogarithms and the symbol map, as well as the construction of single valued multiple polylogarithms and discuss an algorithm for finding fibration bases. We document how these algorithms are implemented in the \mathematica{} package \plt{} and show how it can be used to study the coproduct structure of polylogarithmic expressions and how to compute iterated parametric integrals over polylogarithmic expressions that show up in Feynman integal computations at low loop orders.
}

\preprint{\begin{minipage}[t]{8cm}\begin{flushright}CP3-19-17, CERN-TH-2019-045\\
SLAC-PUB-17423\end{flushright}\end{minipage}}

\begin{document}
\maketitle

\section{Introduction} 
\label{sec:intro}
% !TEX root = polylogtools.tex

Multiple polylogarithms (MPLs) are arguably among the most important class of functions in many areas of modern high-energy physics.  
Beyond the mathematical study of MPLs and their special values, the multiple zeta values (MZVs)~\cite{Goncharov:1996,Goncharov:1998kja,GoncharovMixedTate,Goncharov:2005sla,Brown:2009ta,Brown:2009qja,Brown:2011ik,Brown:2013gia,Gangl1,Gangl2,CharltonDan,CharltonThesis}, these objects have played an important role in many developments in high-energy physics in recent years, ranging from the study of scattering amplitudes in various string theories~\cite{Mafra:2011nw,Schlotterer:2012ny,Broedel:2013tta}, supergravity theories~\cite{Abreu:2019rpt,Chicherin:2019xeg} and maximally supersymmetric Yang-Mills (SYM) theory~\cite{Goncharov:2010jf,Golden:2013xva,Spradlin:2011wp,Dixon:2014voa,Dixon:2014iba,Dixon:2015iva,Dixon:2016nkn,Dixon:2016apl,Caron-Huot:2016owq,Abreu:2018aqd} to the calculation of scattering amplitudes and cross sections for processes at the Large Hadron Collider (LHC)~\cite{Duhr:2012fh,Chavez:2012kn,Anastasiou:2013srw,Anastasiou:2013mca,Dulat:2014mda,Anastasiou:2014vaa,Anastasiou:2014lda,Anastasiou:2015ema,Mistlberger:2018etf,Dulat:2017brz,Dulat:2017prg,Dulat:2018bfe,Chicherin:2018old,Chicherin:2018yne}.
Even before the understanding of multiple polylogarithms as a general class of functions, various special cases have played important roles in the physics literature. The appearance of the dilogarithm one-loop virtual corrections has been known since the early days of QED. In ref.~\cite{Remiddi:1999ew} Vermaseren and Remiddi introduced a special class of multiple polylogarithms, which they dubbed harmonic polylogarithms due to their relation to harmonic sums. These functions have appeared in virtually every multiloop calculation in the last twenty years. In recent years the various special classes of polylogarithms have been understood as arising from the general class of multiple polylogarithms.

An important problem in the practical use of these various classes of functions has been the wealth of functional identities that they satisfy. In many cases these functional identities needed to be derived manually for each case under consideration, hampering an efficient use of the more general classes of polylogarithms.
However, in recent years, Goncharov's seminal paper~\cite{Goncharov:2005sla} on the Hopf algebra structure of multiple polylogarithms has spurred a flurry of research both on the side of mathematics as well as in physics to better understand the algebraic structures of multiple polylogarithms. This has lead in particular to the realization that the Hopf algebra of multiple polylogarithms and its associated coproduct or coaction allow the derivation of functional identities between MPLs in a purely combinatorial fashion, see, e.g., refs.~\cite{Brown:2011ik,Duhr:2012fh}. Based on this, an algorithmic way to compute Feynman integrals was devised~\cite{Brown:2011ik,Anastasiou:2013srw} that makes use of the ability to derive any needed functional equation using the coproduct. Similar algorithms were further developed and automated in ref.~\cite{Ablinger:2014yaa,Panzer:2015ida,Panzer:2014caa,Bogner:2015nda}. In particular, the package {\sc HyperInt} has recently been used to evaluate complicated Feynman parameter integrals~\cite{Bourjaily:2017bsb,Bourjaily:2018aeq,Bourjaily:2018yfy,Bourjaily:2019jrk}.

In parallel the algebraic properties of multiple polylogarithms have been exploited in more formal developments such as the study of motivic amplitudes~\cite{Goncharov:2010jf,Golden:2013xva} and the amplitudes / cluster bootstrap in planar $\mathcal{N}=4$ SYM~\cite{Dixon:2013eka,Dixon:2014voa,Dixon:2014iba,Dixon:2015iva,Caron-Huot:2016owq,Dixon:2016nkn,Drummond:2014ffa,Drummond:2018caf}. These approaches exploit a particular component of the Hopf algebra of the multiple polylogarithms, the so-called symbol map, to study the algebraic structure of scattering amplitudes. Building on the understanding of the branch-cut structure of the multiple polylogarithms, it was also possible to employ the coaction to build a special class of multiple polylogarithms, called single-valued multiple polylogarithms (SVMPLs) that are free of branch cuts~\cite{Brown:2004ugm,Brown:2013gia}. These SVMPLs have played an important role in the study of scattering amplitudes in the so-called multi-Regge Limit~\cite{DelDuca:2016lad,DelDuca:2017peo,DelDuca:2018hrv,Marzucca:2018ydt} and in the calculation and study of certain periods in $\phi^4$-theory~\cite{Schnetz:2013hqa}. Inspired by the Hopf-algebra of the multiple polylogarithms and Brown's generalization to a conjectured cosmic Galois group~\cite{Brown:2015fyf}, it was shown in refs.~\cite{Abreu:2017ptx,Abreu:2017enx,Abreu:2017mtm} that the coaction of one-loop Feynman integrals can be cast in a remarkably simple and compact form. 
%Recently, the study of multiple polylogarithms culminated in a first construction of a elliptic generalizations of multiple polylogarithms that accomodate an elliptic coaction~\cite{Broedel:2017kkb,Broedel:2017siw,Broedel:2018iwv,Broedel:2018qkq}.

All these developments have profited enormously from the mathematical advances in the understanding of the algebraic properties of the multiple polylogarithms. The goal of this paper is to document the \mathematica\ package \plt{} that provides an implementation of many of these algebraic structures. The focus of this implementation is on providing a basis for future exploration and experimentation, as well the ability to compute Feynman integrals that appear in two-loop and three-loop amplitudes~\cite{Anastasiou:2013srw}. This package in particular is not optimized for the performance that would be required to tackle Feynman integrals at arbitrary high loop orders, but it provides well-tested implementations of all required algorithms and provides the flexibility to adapt it to virtually any application mentioned above.\footnote{In fact private versions of this package have been used in many recent computations, see Section~\ref{sec:validation}.} \plt{} provides many routines that allow the user to work with multiple polylogarithms in \mathematica; from integrating expressions in terms of polylogarithms through numerical evaluation and calculation of symbols and coproducts to the automated derivation of functional identities. The goal of this paper is to review the relevant mathematics of polylogarithms and to document the routines in \plt{} that implement the respective mathematical algorithms.

This paper is organized as follows: In section~\ref{sec:installation} we describe how to obtain and install the package and its prerequisites. Then, in section~\ref{sec:MPLs} we review the basic construction of the multiple polylogarithms and describe how they are realized in \plt{}. In section~\ref{sec:shuffle_stuffle} we review the first important algebraic structure that multiple polylogarithms are equipped with, the so-called shuffle algebra, and show how to use it in the package. Next, in section~\ref{sec:special_values}, we discuss the degeneration of multiple polylogarithms to multiple zeta values for special values of their arguments and show how they are implemented in \plt{}. In section~\ref{sec:coproduct} we review the Hopf algebra and coproduct of multiple polylogarithms and its implementation. Then, in section~\ref{sec:symbols} we discuss the relation between the coproduct and the symbol map. In section~\ref{sec:basic} we discuss how to perform basic calculations in \plt{}. In section~\ref{sec:sv_mpls} we review the theory of SVMPLs and discuss the routines dedicated to handling these functions. Afterwards, in section~\ref{sec:validation} we review a few applications of \plt{} in the literature. Finally, in section~\ref{sec:example} we discuss the calculation of a Feynman integral from start to finish to illustrate the practical use of \plt{} for actual calculations.

\section{Installation of the package}
\label{sec:installation}
% !TEX root = polylogtools.tex
\polylogtools{} resides in a git repository at \href{https://gitlab.com/pltteam/plt}{https://gitlab.com/pltteam/plt}. From this gitlab URL it is possible to download a compressed file with the latest version of the repository. Alternatively, \polylogtools{} can be obtained by cloning the repository using git, e.g.~by issuing the following command in any shell that has the \texttt{git} command available:
\begin{verbatim}
    git clone https://gitlab.com/pltteam/plt.git
\end{verbatim}
This will clone the \polylogtools{} repository into a subfolder \texttt{plt}. From within that folder it is then possible to obtain the latest changes and bugfixes to \polylogtools{} by simply issuing the following command on the shell:
\begin{verbatim}
    git pull
\end{verbatim}

\polylogtools\ depends on the \mathematica\ package \hpl\ and the \ginac\ library. Both codes need to be installed separately before \polylogtools\ can be loaded into the current \mathematica\ session. The path in which \hpl\ has been installed needs to be present in the \mathematica\ variable {\tt \$Path}, which should be the case if \hpl\ is installed properly. The \ginac\ library needs to be installed with support for the {\tt ginsh} command line tool and {\tt ginsh} needs to be in a path that can be found by the system (i.e., it must be possible to start \ginac\ by simply typing {\tt ginsh} into a shell command line).
We note that there seems to be a problem in \mathematica\ on MacOS that prevents \mathematica\ from running the {\tt ginsh} command from within a \mathematica\ session unless \mathematica\ has been started from the terminal. This can usually be achieved by running the following command from the terminal:
\begin{verbatim}
/Applications/Mathematica.app/Contents/MacOS/Mathematica
\end{verbatim}

Afterwards \polylogtools\ can be loaded using:
\exbox{
\inline{1}{& {\tt \$PolyLogPath = SetDirectory[ < path > ];}}\\
\inline{2}{& {\tt <<PolyLogTools`;}}
}
where {\tt < path >} is the path to the folder containing the file {\tt PolyLogTools.m}. 
This loads the \mathematica\ packages \hpl\ and {\sc Combinatorica} into the kernel as well. Note that \plt{}  tries to make sure that \texttt{ginsh} is available by running `\texttt{which ginsh}'. On some systems this can fail due to the \texttt{which} command not being available. In these cases \plt{} will complain about not being able to locate \texttt{ginsh}, however, if \texttt{ginsh} is accessible from the terminal, this warning can be safely ignored.

\section{Multiple polylogarithms in \polylogtools}
\label{sec:MPLs}
% !TEX root = polylogtools.tex

In this section we give a short review of the main actors in the \polylogtools\ package, the \textit{multiple polylogarithms} (MPLs). 
The aim of this section is not to provide an extensive overview of MPLs but to introduce our notations and conventions, and how these functions and some of their most basic properties are implemented into \polylogtools.

MPLs can be defined recursively via the iterated integral ($n\geq 0$)~\cite{Goncharov:1998kja,GoncharovMixedTate}
 \beq\label{eq:MPL_def}
 G(a_1,\ldots,a_n;z)=\,\int_0^z\,{d t\over t-a_1}\,G(a_2,\ldots,a_n;t)\,,\\
\eeq
with $G(z) = 1$ and $a_i$ and $z$ are complex variables. We call the vector $\vec a = (a_1,\ldots,a_n)$ the \textit{weight vector}, and its length $n$ is called the \textit{weight}. In the special case where all the $a_i$'s are zero, we define, using the obvious vector notation $\vec a_n=(\underbrace{a,\dots,a}_{n})$,
\beq\label{eq:G_0_def}
G(\vec 0_n;z) = {1\over n!}\,\log^n z\,,
\eeq
consistent with the case $n=0$ above. In the case where the $a_i$ are constants, MPLs are often referred to as hyperlogarithms.  
MPLs define a very general class of functions that generalize the well-known logarithm and (Nielsen) polylogarithm functions, e.g., for $a\neq 0$,
\beq\bsp
G(\vec a_n;z) &\,= \frac{1}{n!}\log^n\left(1-\frac{z}{a}\right)\,, \\
G(\vec 0_{n-1},a;z) &\,= -\li{n}\left(\frac{z}{a}\right)\,,\\
G(\vec 0_{n-k},\vec a_k;z) &\,= (-1)^k\,S_{n-k,k}\left(\frac{z}{a}\right)\,.
\esp\eeq

The function $G(a_1,\ldots,a_n;z)$ is represented within \polylogtools\ by the symbol {\tt G[a1,\ldots,an,z]}. Here the arguments {\tt a1},\ldots,{\tt an} and {\tt z} can be any valid \mathematica\ expression (in practice, we will only deal with cases where the arguments are rational or algebraic expressions). Since $G(z)=1$, the function {\tt G} automatically evaluates to unity if it only has a single argument.

In the case where all the $a_i$ are 0 or $\pm1$, MPLs are often referred to as \textit{harmonic polylogarithms} (HPLs) in the physics literature~\cite{Remiddi:1999ew}. HPLs are equal to MPLs, up to a sign,
\beq
H(a_1,\ldots,a_n;z) = (-1)^p\,G(a_1,\ldots,a_n;z)\,,\quad a_i\in\{0,\pm1\}\,,
\eeq
where $p$ denotes the number of elements in $\vec a$ equal to $(+1)$. HPLs are represented within \polylogtools\ by the symbols {\tt H[a1,\ldots,an,z]}. 

The functions {\tt HToG[expr]} and {\tt GToH[expr]} allow one to switch from the {\tt H} to the {\tt G} notation inside the \mathematica\ expression {\tt expr}. Note that since the \hpl\ package is loaded automatically with \polylogtools, also the notation for HPLs from the \hpl\ package can be used inside \polylogtools. The functions {\tt HToHPL[expr]}, {\tt GToHPL[expr]}, {\tt HPLToH[expr]} and {\tt HPLToG[expr]} allow the user to switch between the notations used by \hpl\ and \polylogtools.
In particular, the function {\tt HPLToG[expr]} can be used to convert a HPL in compressed notation to a {\tt G} in standard notation, for example:
\exbox{
\inline{1}{& {\tt HPLToG[ HPL[\{2\},x] ]}}\\
\breakline\\
\outline{1}{& {\tt -G[0,1,x] }}
  }

There is a second way to define MPLs, using nested sums rather than iterated integrals~\cite{Goncharov:1998kja}:
\begin{eqnarray}\label{eq:Li_m_def}
\textrm{Li}_{m_1,\ldots,m_k}(z_1,\ldots,z_k) &=& \sum_{0<n_1<n_2<\dots <n_k} \frac{z_1^{n_1} z_2^{n_2} \cdots z_k^{n_k} }{n_1^{m_1} n_2^{m_2} \cdots n_k^{m_k} }\\
& =&
 \sum_{n_k=1}^\infty{z_k^{n_k}\over n_k^{m_k}}\,\sum_{n_{k-1}=1}^{n_k-1}\ldots\sum_{n_1=1}^{n_{2}-1}{z_1^{n_1}\over n_1^{m_1}}\,,\nonumber
\end{eqnarray}
where this definition makes sense whenever the sums converge (e.g., for $|z_i|<1$). The number $k$ of indices is called the \textit{depth} of the MPL. 
The function $\textrm{Li}_{m_1,\ldots,m_k}(z_1,\ldots,z_k)$ is represented inside \polylogtools\ by the symbol {\tt Li[\{m1,\ldots,mk\}, \{z1,\ldots,zk\}]}. 
For depth $k=1$ this definition naturally reduces to the usual series representation of the polylogarithm function $\li{m}(z)$.
The function {\tt Li[\{m\}, \{z\}]} is therefore equivalent to the built-in \mathematica\ function {\tt PolyLog[m,z]}.
The $G$ and Li functions define (essentially) the same class of functions and are related by ($a_i\neq 0$)
\beq\bsp\label{eq:Gm_def}
%\textrm{Li}&_{m_1,\ldots,m_k}(z_1,\ldots,z_k) 
%= (-1)^k\,G\Big(\vec{0}_{m_k-1},{1\over z_k}, \ldots,\vec{0}_{m_1-1},{1\over z_1\ldots z_k};1\Big)\,.
G\Big(\vec{0}_{m_1-1},a_1, \ldots,\vec{0}_{m_k-1},a_k;z\Big) = (-1)^k\,
\textrm{Li}&_{m_k,\ldots,m_1}\left(\frac{a_{k-1}}{a_k},\ldots,\frac{a_1}{a_2},\frac{z}{a_1}\right)\,.
\esp\eeq
The functions {\tt LiToG[expr]} and {\tt GToLi[expr]} allow the user to switch between the two representations of MPLs, either as iterated integrals ({\tt G}) or as nested sums ({\tt Li}).

\begin{table}
\bgfb
\multicolumn{2}{c}{\textbf{Table~\ref{tab:basic}: Definition of MPLs}}\\
\multicolumn{2}{l}{In addition to the functions defined by \hpl:}
\\
\tt{G[a1,a2,\dots,an,z]} & The multiple polylogarithm $G(a_1,\ldots,a_n;z)$. The indices  $a_i$ can be numbers, symbolic constants or functions of other variables.
\\
\tt{H[a1,a2,\dots,an,x]} & The harmonic polylogarithm $H(a_1,\ldots,a_n;z)$. The indices $a_i$ have to be integers from the set $\{-1,0,1\}$.
\\
\tt{Li[\{m1,\dots,mn\},\{z1,\dots,zn\}]} & The multiple polylogarithm $\textnormal{Li}_{m_1,\ldots,m_n}(z_1,\ldots,z_n)$ in sum notation. The $m_i$ are integers.
\\
\multicolumn{2}{l}{Conversion functions:}
\\ 
\tt{GToH[expr]} & Replaces every {\tt G} in the expression {\tt expr} with the corresponding {\tt H}, if the indices of the multiple polylogarithm are integers from the set $\{-1,0,1\}$.
\\
\tt{HToHPL[expr]} & Converts every {\tt H} in {\tt expr} to the {\tt HPL} notation.
\\
\tt{GToHPL[expr]} & Replaces every {\tt G} in {\tt expr} with the corresponding {\tt HPL}, if the indices of the multiple polylogarithm are integers from the set $\{-1,0,1\}$.
\\
\tt{HPLToH[expr]} & Replaces every {\tt HPL} in {\tt expr} with the corresponding {\tt H}. Automatically converts from the $a-$notation defined by \hpl\ to the $m-$notation.
\\
\tt{HToG[expr]} & Replaces every {\tt H} in {\tt expr} with the corresponding {\tt G}.
\\
\tt{HPLToG[expr]} & Replaces every {\tt HPL} in {\tt expr} with the corresponding {\tt G}. Automatically converts from the $a-$notation defined by \hpl\ to the $m-$notation.
\\
\tt{GToLi[expr]} & Converts every {\tt G} in {\tt  expr} to the {\tt Li} notation. 
%Automatically, extracts logarithmic divergences with {\tt ExtractZeroes} to make sure that all multiple polylogarithms are well defined.
\\
\tt{LiToG[expr]} & Converts every {\tt Li} and classical polylogarithm in standard \mathematica\ notation in {\tt expr} to the {\tt G} notation.
\egfb
\textcolor{white}{\caption{\label{tab:basic}}}
\end{table}

\section{The shuffle and stuffle algebras of MPLs}
\label{sec:shuffle_stuffle}
% !TEX root = polylogtools.tex

One of the most basic properties of MPLs is that they can be equipped with two algebra structures, one related to the representation of MPLs as iterated integrals -- the shuffle algebra -- and another one related to the representation as nested series -- the stuffle algebra. Most of the functionalities of \polylogtools\ rely on the shuffle algebra properties of MPLs. Since we will encounter another shuffle algebra in Section~\ref{sec:symbols}, we start this section by reviewing the mathematics of shuffle algebras in general before discussing the implementation of the shuffle algebra of MPLs in \polylogtools.

\subsection{Shuffle algebras}
\label{sec:shuffle_algebras}
Consider a (finite) set $A$, whose elements we will denote as \textit{letters}, and we call $A$ the \textit{alphabet}. For concreteness, we will choose here $A= \{a,b,c,\ldots\}$, though $A$ can be any finite set. We define a vector space $V$ as the vector space formed by all $\mathbb{Q}$-linear combinations of words formed from the letters in $A$, including the empty word, which we denote simply as $1$. The \textit{length} of a word is defined as the number of letters that the word is made of. Words of length 1 are simply the letters. The concatenation of two words $w_1$ and $w_2$ is defined in the obvious way and denoted by $w_1w_2$. $V$ has the structure of a \textit{graded} vector space, i.e., it admits a direct sum decomposition
\beq
V = \bigoplus_{n=0}^{\infty}V_n = V_0 \oplus V_{>0}\,,
\eeq
where $V_0=\mathbb{Q}$ and $V_n$ and $V_{>0}$ denote the subspaces of $V$ spanned by all words of length $n$ and all words of non-zero length respectively.

$V$ can be given the structure of a commutative algebra equipped with the \textit{shuffle product}. The shuffle product assigns to a pair of words $(w_1,w_2)$ the sum of all their \textit{shuffles}, i.e., the sum of all possible ways of permuting the letters of their union without changing the order of the letters within each word. The shuffle product can also be defined recursively: if $\alpha,\beta\in A$ are letters and $w_1,w_2\in V$ are words, then the shuffle product of the words $\alpha w_1$ and $\beta w_2$ is defined recursively by
\beq
\alpha w_1\shuffle \beta w_2 = \alpha(w_1\shuffle \beta w_2) + \beta(\alpha w_1\shuffle w_2)\,,
\eeq
and the recursion starts with $1\shuffle w=w\shuffle 1=w$. For example, the shuffle product of the words $ab$ and $cd$ is
\beq\label{eq:shuffle_word_example}
ab\shuffle cd = abcd + acbd+cabd + acdb+cadb+cdab\,.
\eeq
It is easy to check that the shuffle product is associative and commutative. Moreover, it preserves the length,  i.e., the shuffle product of two words of lengths $n_1$ and $n_2$ is a linear combination of words of length $n_1+n_2$. In this way $V$ becomes a \textit{graded} algebra,
\beq
V_{n_1}\shuffle V_{n_2} \subseteq V_{n_1+n_2}\,.
\eeq

It is often convenient to work with a set of generators for $V$, i.e., a minimal set of words such that every element of $V$ can be written as a linear combination of products (a polynomial) in these generators. In order to define such a set of generators, we first need to define an ordering $<$ on the set of letters $A$ (for concreteness, we can choose here the lexicographic ordering among the letters, but any other choice of ordering would do). A theorem by Radford states that the shuffle algebra $V$ is isomorphic to the polynomial algebra (over $\mathbb{Q}$) formed by a subset of words, called \textit{Lyndon words}~\cite{Radford}.
A Lyndon word is a non-empty word that is less (for the chosen ordering $<$) than any of its proper right factors, i.e., a word $w$ is a Lyndon word if for all non-empty words $w_1$ and $w_2$ with $w=w_1w_2$ we have $w< w_2$.
A consequence of Radford's theorem is that for a given ordering every word can be uniquely written as a polynomial in Lyndon words. For example, the word $baa$ is not a Lyndon word for the lexicographic ordering (because $baa > aa$), so it can be written as a polynomial in the Lyndon words,
\beq\label{eq:Lyndon_example}
baa = aab - a\shuffle ab + \frac{1}{2}\,a\shuffle a\shuffle b\,.
\eeq

\subsection{The shuffle algebra of MPLs}
\label{sec:shuffle_MPLs}

Let us now consider the $\mathbb{Q}$-vector space of all MPLs of the form $G(\vec a;z)$ ending in the same variable $z$. It can be shown that this vector space forms a graded shuffle algebra, where $G(\vec a;z)$ is identified with the word $\vec a = (a_1,\ldots, a_n)$. The weight of an MPL corresponds to the length of the word. Explicitly, the shuffle product of two MPLs ending in the same variable can be written as
\beq
G(\vec a;z)\,G(\vec b;z) = \sum_{\vec c = \vec a\shuffle \vec b} G(\vec c;z)\,,
\eeq
where the sum runs over all shuffles of the two weight vectors $\vec a$ and $\vec b$.

It is possible to use \polylogtools\ to linearise all shuffle products of MPLs ending in the same argument $z$ by using the function {\tt ShuffleG} as shown in the following example (cf.~eq.~\eqref{eq:shuffle_word_example}),
\exbox{
\inline{1}{& {\tt ShuffleG[ G[a,b,z] * G[c,d,z] ]}}\\
\breakline\\
\outline{1}{& {\tt G[a,b,c,d,z] + G[a,c,b,d,z] + G[a,c,d,b,z] + G[c,a,b,d,z] +} \\ 
&{\tt  G[c,a,d,b,z] + G[c,d,a,b,z]}}
  }

Since every word can be represented as a polynomial in Lyndon words for a given ordering, we can also decompose every MPL into a linear combination of products of MPLs whose weight vectors are Lyndon words. This is achieved as follows (cf.~eq.~\eqref{eq:Lyndon_example}), 
\exbox{
\inline{2}{& {\tt DecomposeToLyndonWords[ G[b,a,a,z], Alphabet -> \{a,b\} ]}}\\
\breakline\\
\outline{2}{& {\tt G[a,a,b,z] - G[a,z] * G[a,b,z]  + {1/2} * G[a,z]\^{}2 * G[b,z]}}
  }
The function {\tt DecomposeToLyndonWords} does not only work on single MPLs, but its argument can be any \mathematica\ expression. \polylogtools\ has tables of Lyndon words hard-coded up to words of length six, and so only MPLs up to weight six are reduced to Lyndon words. The tables of Lyndon words implemented in \polylogtools\ have been generated with {\sc Sage}~\cite{sage}. The optional argument {\tt  Alphabet} takes as value a list of symbols and defines the alphabet as well as the ordering among the letters (simply the order in which the letters appear inside the list). Its default value is {\tt \{0,1,-1\}}. Note that only those MPLs whose weight vectors contain only letters from the {\tt Alphabet} are decomposed into Lyndon words. Finally, let us make a comment about singularities. The MPL $G(a_1,\ldots,a_n;z)$ diverges whenever $z=a_1$, because in that case the integral in eq.~\eqref{eq:MPL_def} has an end-point singularity. It may happen that the decomposition into Lyndon words leads to divergent MPLs, even though the input was finite (e.g., take {\tt z=a} in the above example box). In such a case the decomposition is not performed and a warning is shown.

It is often convenient to decompose MPLs not into a Lyndon word basis, but into a set of functions corresponding to words where a given letter does not appear in the last position (except for words that are only made of that letter). For example, in applications it is useful to represent a function in terms of MPLs where the last letter is non-zero, except for powers of $G(0;z)$. Such a representation can always be achieved algorithmically by unshuffling powers of logarithms. For example, we have
\beq
G(1,0,0;z) = G(0,0,1;z) - G(0;z)\,G(0,1;z) + G(0,0;z)G(1;z)\,.
\eeq
This operation is implemented via the function {\tt ExtractZeroes}, which can be used as shown,
\exbox{
\inline{3}{& {\tt ExtractZeroes[ G[1,0,0,z] ]} }\\
\breakline\\
\outline{3}{& {\tt G[0,0,1,z] - G[0,z]*G[0,1,z]  + G[0,0,z]*G[1,z]}}
  }

%%%%%%%%%%%%%%%%%%%%%%%%%%%%%%%%%%%%%%%%%%%%%
\subsection{The stuffle algebra of MPLs}

There is another algebra structure on MPLs coming from their series representation, called \textit{stuffle algebra}~\cite{Hoffman:1999}. Since in many applications to Feynman integrals the shuffle algebra structure seems more relevant, we will be brief and only present an example. For a more general discussion, we refer to ref.~\cite{Hoffman:1999}. Consider a product of two polylogarithms of depth one. We can rearrange the sums in the following way:
\beq\bsp\label{eq:stuffle_example}
\li{m_1}(z_1)\,\li{m_2}(z_2) &\,= \left(\sum_{n_1=1}^\infty\frac{z_1^{n_1}}{n_1^{m_1}}\right)\,\left(\sum_{n_2=1}^\infty\frac{z_2^{n_2}}{n_2^{m_2}}\right)=\sum_{n_1,n_2=1}^{\infty}\frac{z_1^{n_1}z_2^{n_2}}{n_1^{m_1}n_2^{m_2}}\\
&\,=\sum_{\substack{n_1=1\\n_2<n_1}}^{\infty}\frac{z_1^{n_1}z_2^{n_2}}{n_1^{m_1}n_2^{m_2}}+\sum_{\substack{n_2=1\\n_1<n_2}}^{\infty}\frac{z_1^{n_1}z_2^{n_2}}{n_1^{m_1}n_2^{m_2}}+\sum_{n=1}^{\infty}\frac{(z_1z_2)^{n}}{n^{m_1+m_2}}\\
&\,=\li{m_1,m_2}(z_1,z_2) + \li{m_2,m_1}(z_2,z_1) + \li{m_1+m_2}(z_1\,z_2)\,.
\esp\eeq
Similar formulas can be derived for stuffle products of MPLs of higher depths. \polylogtools\ can also expand products of Li-functions into stuffle products. For example, the stuffle product in eq.~\eqref{eq:stuffle_example} is obtained as follows,
\exbox{
\inline{1}{& {\tt StuffleLi[ Li[\{m1\},\{z1\}] * Li[\{m2\},\{z2\}] ]}}\\
\breakline\\
\outline{1}{& {\tt Li[\{m1,m2\},\{z1,z2\}]+Li[\{m2,m1\},\{z2,z1\}]+Li[\{m1+m2\},\{z1*z2\}]}}
  }

%%%%%%%%%%%%%%%%%%%%%%%%%%%%%%%%%%%%%%%%%%%%

\begin{table}
\bgfb
\multicolumn{2}{c}{\textbf{Table~\ref{tab:shuffle}: Shuffle \& stuffle products}}
\\
{\tt ShuffleG[expr]}& Expands all products of $G$-functions that end in the same argument into a sum of shuffles.
\\
{\tt StuffleLi[expr]}& Expands all products of Li-functions into a sum of stuffles.
\\
{\tt DecomposeToLyndonWords[expr]}& Decomposes all $G$-functions into a sum of products of Lyndon words. The set of letters and their ordering can be passed as a list via the optional argument {\tt Alphabet}. The default value is {\tt {\tt \{0,1,-1\}}}. MPLs which depend on a letter that is not in the {\tt Alphabet} are not decomposed. The same applies to MPLs that would be decomposed into divergent quantities.
\\
{\tt ExtractZeroes[expr]}& Uses the shuffle algebra to remove all trailing zeroes from all {\tt G}-functions in {\tt expr}.
\egfb
\textcolor{white}{\caption{\label{tab:shuffle}}}
\end{table}

\section{Special values of MPLs}
\label{sec:special_values}
% !TEX root = polylogtools.tex

When evaluated at certain special values of the arguments, MPLs often reduce to known transcendental constants. In particular, if HPLs are evaluated at $z=\pm1$, then they reduce to (coloured) multiple zeta values (MZVs),
\beq\bsp\label{eq:mzv_def}
\zeta&_{m_k,\ldots,m_1}
= \sum_{0<n_1<n_2<\dots <n_k} \frac{s_1^{n_1} s_2^{n_2} \cdots s_k^{n_k} }{n_1^{|m_1|} n_2^{|m_2|} \cdots n_k^{|m_k|} }\,,\quad s_i = \textrm{sign}(m_i)\,.
\esp\eeq
The weight and the depth of (coloured) multiple zeta values are defined in the same way as for MPLs. The series in eq.~\eqref{eq:mzv_def} diverges whenever $m_k=1$. For depth $k=1$, MZVs reduce to Riemann's zeta function at positive integer arguments,
\beq
\zeta_m = \zeta(m) = \sum_{n=1}^\infty \frac{1}{n^m}\,.
\eeq

MZVs and their coloured generalisations (which correspond to some of the $m_i$ being negative) are implemented in the \hpl\ package, where $\zeta_{m_k,\ldots,m_1}$ is represented by the symbol {\tt MZV[\{mk,\ldots,m1\}]}. In addition, the \hpl\ package also knows how to reduce HPLs evaluated at $z=\pm i$ to a smaller set of transcendental constants. Since \polylogtools\ uses \hpl\, the reduction rules of HPLs evaluated at $z\in\{\pm 1,\pm i\}$ are readily available in \polylogtools\ also for the {\tt G} and {\tt H} functions. We refer to the documentation of the \hpl\ package for further details~\cite{Maitre:2005uu,Maitre:2007kp}.
In addition to the values of HPLs evaluated at $z=\pm i$, \polylogtools\ can also reduce $G(\vec a;z)$ with $a_i\in\{0,\pm i\}$ and $z\in \{\pm 1,\pm i\}$, as well as HPLs for some small weight evaluated at $z\in\{\pm 1/2,\pm 1/3\}$, to known transcendental constants. 

By default, \polylogtools\ will automatically express any MPL in terms of transcendental constants whenever possible. It is possible to disable the automatic reduction to transcendental constants by setting 
\exbox{
\inline{1}{& {\tt PLT\$AutoConvertToKnownConstants = False;}}
  }
  The default value of {\tt PLT\$AutoConvertToKnownConstants} is {\tt True}.
  
  We conclude this section with a comment on the regularisation of MPLs. We have seen that $G(a_1,\ldots,a_n;z)$ diverges whenever $z=a_1$. Hence, when evaluating HPLs at $z=\pm1$ the result may be divergent. It is easy to see that all divergencies are logarithmic, and it is useful to introduce a formal quantity $\lim_{z\to 1}G(1;z)=-\zeta_1$ which acts as a regulator and allows one to keep track of divergences. Indeed, all divergencies show up as powers of $\zeta_1$, and the dependence on the regulator $\zeta_1$ must cancel for finite quantities. The regulator $\zeta_1$ is represented by {\tt HPL[\{1\},1]} in \hpl\ and \polylogtools.
  
  There is, however, a subtle point about the regularisation discussed in the previous paragraph. As usual, there is an ambiguity in how to introduce the regulator, related to shifting divergent quantities by finite terms. In practice, one would like to choose a regulator that preserves as many of the algebraic structures as possible. MPLs are equipped with both a shuffle and a stuffle algebra structure. However, it is not possible to preserve both structures at the same time, leading to two different `schemes' to regulate MPLs and MZVs~\cite{ShuffleReg}, referred to as \textit{shuffle regularisation} and \textit{stuffle regularisation} respectively. For example, using the shuffle product on MPLs, we obtain
  \beq
  \zeta_1^2 = \lim_{z\to 1} G(1;z)^2 = \lim_{z\to1}2\,G(1,1;z)\,,
  \eeq 
  so that the shuffle-regulated value of $G(1,1;z)$ at $z=1$ is given by $\frac{1}{2}\zeta_1^2$.
  Instead, using the fact that $G(1;z) = -\li{1}(z)$ as well as the stuffle product, we find
  \beq\bsp
  \zeta_1^2 &\,= \lim_{z\to1}G(1;z)^2 = \lim_{z\to 1}\li{1}(z)^2 = \lim_{z\to 1} 2\li{1,1}(z,z) +\li{2}(z^2)\\
&\,  = \zeta_2 + \lim_{z\to 1} 2\,G(1,1;z)\,,
  \esp\eeq
 so that the stuffle-regulated value of $G(1,1;z)$ at $z=1$ is given by $\frac{1}{2}\zeta_1^2-\frac{1}{2}\zeta_2$.
 We see that the shuffle- and stuffle-regulated values are different. Note that any finite quantity must be independent of the regularisation scheme chosen to compute it. However, care is needed that the same scheme is applied consistently throughout a computation!
 
 By default, \polylogtools\ uses shuffle regularisation to regulate the {\tt G} and {\tt H} functions, because in applications it is often more convenient to preserve the shuffle algebra structure (which is more closely related to the definition of MPLs as iterated integrals). It is possible to switch to stuffle regularisation by changing the value of the variable {\tt PLT\$ShuffleRegularisation} from {\tt True} to {\tt False}. We point out that, unlike \polylogtools, the \hpl\ package uses stuffle regularisation to define the values of MZVs and HPLs at $z=\pm1$. Care is thus needed if the output and functions from this package are used in conjunction with \polylogtools.

\section{The Hopf algebra of MPLs}
\label{sec:coproduct}
% !TEX root = polylogtools.tex

An important property of MPLs is that they can be equipped with a coproduct turning them into a Hopf algebra. The Hopf algebra of MPLs has led to the development of novel techniques to deal with Feynman integrals that evaluate to MPLs, and it has seen a multitude of applications over the last few years. In particular, it allows one to control more rigorously identities among MPLs, see ref.~\cite{Duhr:2012fh,Duhr:2014woa} for a pedagogical introduction of how to derive identities among MPLs using the coproduct. Another important application of the coproduct is the amplitude bootstrap in planar $\mathcal{N}=4$ Super Yang-Mills~\cite{Dixon:2013eka,Dixon:2014iba,Dixon:2014voa,Dixon:2015iva,Caron-Huot:2016owq,Dixon:2016nkn}. Here we will not discuss these applications in detail, but we focus on the implementation of the coproduct on MPLs in \polylogtools. Since we will discuss different Hopf algebras in subsequent sections, we give a short review of Hopf algebras in general before focusing on the case of MPLs.

\subsection{A short review of Hopf algebras}
\label{sec:Hopf_review}

A \textit{coalgebra} is a $\mathbb{Q}$-vector space $H$ together with a linear map $\Delta:H\to H\otimes H$ called the coproduct. The coproduct is required to be \textit{co-associative}, that means 
\beq
(\textrm{id}\otimes\Delta)\Delta = (\Delta\otimes \textrm{id})\Delta\,.
\eeq
A coalgebra must admit another map, called the \textit{counit}, which does not play any role in the context of the Hopf algebras discussed in this paper.
A Hopf algebra is a vector space that is both an algebra and coalgebra, such that the product and the coproduct are compatible,
\beq
\Delta(a\cdot b) = \Delta(a)\cdot \Delta(b)\,.
\eeq
In the previous equation the multiplication of tensors on the right-hand side is defined component-wise,
\beq\label{eq:tensor_mult}
(a_1\otimes a_2)\cdot (b_1\otimes b_2) = (a_1\cdot b_1)\otimes (a_2\cdot b_2)\,.
\eeq
In addition, there is a map $S:H\to H$, called the \textit{antipode}, which will be defined below.

The coalgebras encountered in this paper are \textit{graded} and \textit{connected}, i.e., they admit a direct sum decomposition
\beq
H= \bigoplus_{n=0}^\infty H_n = H_0 \oplus H_{>0}\,,\qquad H_0=\mathbb{Q}\,,
\eeq
and the coproduct respects the grading,
\beq
\Delta(H_n) \subseteq \bigoplus_{p+q=n}H_p\otimes H_q\,.
\eeq

We can iterate the coproduct to obtain tensors with more and more factors. This iteration can be done in different ways, e.g., by iterating on the first or on the second factor of the coproduct. The co-associativity of the coproduct ensures that the different ways of iterating the decomposition into simpler objects give the same result. 
Since the coproduct respects the grading, it makes sense to define $\Delta_{i_1,\ldots,i_k}$ as the the part of the iterated coproduct that takes values in $H_{i_1}\otimes\cdots\otimes H_{i_k}$. The maps $\Delta_{i_1,\ldots,i_k}:H\to H_{i_1}\otimes\cdots\otimes H_{i_k}$ satisfy the obvious recursion
\beq\label{eq:iterated_Delta}
\Delta_{i_1,\ldots,i_k} = (\Delta_{i_1,\ldots,i_{k-1}}\otimes \textrm{id})\Delta_{i,i_k}\,,\qquad i=i_1+\ldots+i_{k-1}\,.
\eeq

Let us also discuss the antipode $S$. In the case of graded and connected Hopf algebras the antipode is uniquely determined by the coproduct to be
\beq \label{eq:antipode_from_delta}
S(x) = -x -m(\textrm{id}\otimes S)\Delta'(x) = -x -m(S\otimes\textrm{id})\Delta'(x)\,,\qquad x\in H_{>0}\,,
\eeq
where $\Delta'(x) = \Delta(x)-1\otimes x-x\otimes 1$ denotes the \textit{reduced coproduct} and $m(a\otimes b) = a\cdot b$ is the multiplication in $H$. The antipode is linear, acts trivially on elements of weight 0, $S(1)=1$, and preserves the multiplication and the coproduct
\beq
S(a\cdot b) = S(b)\cdot S(a) \textrm{~~~and~~~} \Delta S = (S\otimes S)\Delta.
\eeq

Let us conclude this section by illustrating these definitions on a concrete example of a Hopf algebra. This example will come back in Section~\ref{sec:symbols} in the context of MPLs. We start from the shuffle algebra $V$ of words from Section~\ref{sec:shuffle_algebras}. Every shuffle algebra can be turned into a graded and connected Hopf algebra whose coproduct is given by the deconcatenation of words,
\beq\label{eq:Delta_dec}
\Delta_{\textrm{dec}}(w) = \sum_{w=w_1w_2}w_1\otimes w_2\,,
\eeq
where the sum runs over all deconcatenations of the word $w$, and including the trivial ones where either $w_1$ or $w_2$ is the empty word. One can check that this definition satisfies all the properties of a coproduct. The antipode is given by the reversal of words,
\beq\label{eq:antipode_def}
S_{\textrm{dec}}(w) = (-1)^{|w|}\,\tilde{w}\,,
\eeq 
where $|w|$ denotes the length of $w$, and the tilde denotes the reversal of words (e.g., if $w=abc$, then $\tilde{w}=cba$).

%%%%%%%%%%%%%%%%%%%%%%%%%%%%%

\subsection{The Hopf algebra of MPLs}
\label{sec:hopf_mpl}
In ref.~\cite{Goncharov:2005sla} it was argued that MPLs form a graded and connected Hopf algebra, where the grading comes again from the weight of the functions. The explicit formula for the coproduct is rather involved, and most conveniently expressed not in terms of the MPLs defined in eq.~\eqref{eq:MPL_def}, but in terms of the iterated integrals
 \beq\label{eq:MPL_I_def}
 I(a_0;a_1,\ldots,a_n;a_{n+1})=\,\int_{a_0}^{a_{n+1}}\,{d t\over t-a_n}\,I(a_0;a_1,\ldots,a_{n-1};t)\,.
\eeq
It is clear that the functions defined in eqs.~\eqref{eq:MPL_def} and~\eqref{eq:MPL_I_def} define the same space of functions. In terms of the functions defined in eq.~\eqref{eq:MPL_I_def}, the coproduct on MPLs can be written in the following compact form~\cite{Goncharov:2005sla},
\begin{align}\label{eq:coproduct}
\Delta(I(a_0;a_1,\ldots,a_n;a_{n+1})) &\,= \sum_{0=i_1<i_2<\ldots<i_{k}<i_{k+1}=n}\!\!\! I(a_0;a_{i_1},\ldots,a_{i_k};a_{n+1})\\
\nonumber&\qquad \otimes\Bigg[\prod_{p=0}^kI(a_{i_p};a_{i_p+1},\ldots,a_{i_{p+1}-1};a_{i_{p+1}})\Bigg]\,.
\end{align}
Here the sum runs over all ordered subsets of a given length $k$ of $(a_1,\ldots,a_n)$. Since the Hopf algebra of MPLs is graded and connected, the antipode is uniquely determined by the coproduct and eq.~\eqref{eq:antipode_from_delta}, so we do not show it here explicitly.

Let us make some comments about the formula for the coproduct in eq.~\eqref{eq:coproduct}. First, the individual terms in eq.~\eqref{eq:coproduct} admit a simple combinatorial interpretation in terms of polygons inscribed into a semi-circle~\cite{Goncharov:2005sla,GoncharovMixedTate}. It is based on this combinatorial framework that the coproduct on MPLs is implemented in \polylogtools, and not via the explicit formula in eq.~\eqref{eq:coproduct}. We do not review the combinatorial picture based on semi-circles here, but we refer to ref.~\cite{Duhr:2014woa} where worked-out examples can be found. Second, we note that the formula for the coproduct in eq.~\eqref{eq:coproduct} is only valid in the generic case where all the arguments $a_i$ are distinct. In the non-generic case individual terms in eq.~\eqref{eq:coproduct} may diverge, and we need to replace the MPLs on the right-hand side of eq.~\eqref{eq:coproduct} with suitably regularised versions. We follow closely refs.~\cite{Goncharov:2005sla,Duhr:2012fh,Duhr:2014woa}, identifying all MPLs on the right-hand side of eq.~\eqref{eq:coproduct} with their shuffle-regularised versions (cf.~Section~\ref{sec:special_values}), and setting to zero the regulator $\zeta_1$. 

The implementation of the coproduct on MPLs is one of the main features of the \polylogtools\ package, because the coproduct is the basis for many applications. The coproduct is called via the function {\tt Delta}, which takes as argument any valid expression in terms of MPLs of weight up to twelve and returns its coproduct. For example, we have
\exbox{
\inline{1}{& {\tt Delta[ G[a,b,z] ]}}\\
\breakline\\
\outline{1}{& {\tt CT[1, G[a,b,z]] +  CT[G[a,b,z], 1] + CT[G[a,z], G[b,a]] - }\\
& {\tt  CT[G[b,z], G[a,b]] + CT[G[b,z], G[a,z]]}}
}
Tensors are represented in \polylogtools\ by lists with head {\tt CT}, i.e., the
tensor $A_1\otimes A_2\otimes A_3\otimes\ldots$ is represented by the symbol
{\tt CT[A1,A2,A3,...]}.\footnote{{\tt CT} is the abbreviation for {\tt CircleTimes}, which is the name given by \mathematica\ to the $\otimes$ symbol. Note that in \texttt{StandardForm} or \texttt{TraditionalForm} the head \texttt{CT} is automatically formatted to look like a tensor product.} 
Note that {\tt Delta} can be applied to any valid expression made of polylogarithms, independently of how they are represented ({\tt G}, {\tt H}, {\tt Li}, {\tt Log}, {\tt PolyLog}). The only restriction is that no MPLs of weight higher than twelve are allowed in the current implementation. The coproduct also acts non trivially on MZVs (because they are just special values of MPLs), e.g.,
\exbox{
\inline{2}{& {\tt Delta[ MZV[\{5,3\}] ]}}\\
\breakline\\
\outline{2}{& {\tt CT[1,MZV[\{5,3\}]] +  CT[MZV[\{5,3\}],1] - 5*CT[Zeta[3],Zeta[5]]}}
}

The symbol {\tt CT} satisfies the basic properties of a tensor product. In particular, it satisfies $A\otimes (B\otimes C)\otimes D = A\otimes B\otimes C\otimes D $, i.e., we have
\exbox{
\inline{3}{& {\tt CT[A,CT[B,C],D]}}\\
\breakline\\
\outline{3}{& {\tt CT[A,B,C,D]}}
}
Moreover, it is linear with respect to rational numbers and the imaginary unit $i$, e.g.,
\exbox{
\inline{4}{& {\tt CT[A,2*I*G[a,z] + 1/2*G[b,z],B]}}\\
\breakline\\
\outline{4}{& {\tt 2*I*CT[A,G[a,z],B] + 1/2*CT[A,G[b,z],B]}}
}
It also automatically evaluates products of tensors (cf.~eq.~\eqref{eq:tensor_mult}),
\exbox{
\inline{5}{& {\tt CT[a1,b1] * CT[a2,b2]}}\\
\breakline\\
\outline{5}{& {\tt CT[a1*a2,b1*b2]}}
}

The different components $\Delta_{i_1,\ldots,i_k}$ of the iterated coproduct can be accessed as in the following example for $\Delta_{2,1,1}$,
\exbox{
\inline{6}{& {\tt Delta[\{2,1,1\},G[0,0,1,1,z]]}}\\
\breakline\\
\outline{6}{& {\tt CT[G[1,1,z],G[0,z],G[0,z]]}}
}
The different components of the iterated coproduct are evaluated internally using the recursion in eq.~\eqref{eq:iterated_Delta}.

At this point we have to make an important comment: MPLs are multivalued functions. It is only MPLs modulo their discontinuities that form a Hopf algebra. The discontinuities are related to taking residues at the simple poles in the integrand in eq.~\eqref{eq:MPL_def}. Therefore the discontinuities of MPLs are always proportional to $i \pi$. Hence, in order to obtain a Hopf algebra, we need to work modulo $i\pi$. In applications, however, it is important to keep track of powers of $\pi$. It is possible to incorporate $i\pi$ into the construction by introducing the special rule~\cite{Brown:2011ik} (see also ref.~\cite{Duhr:2012fh}),\footnote{This special rule can be motivated because one obtains a co-module equipped with a coaction. By abuse of language, we will only refer here to the Hopf algebra and the coproduct, as in physics applications the distinction is often minor. Internally \plt{} always computes the coaction.}
\beq
\Delta(i \pi) = i\pi \otimes1\,.
\eeq
More precisely, we should work modulo $i\pi$ in all entries of the coproduct apart from the first. For example, in \polylogtools\ one obtains
\exbox{
\inline{7}{& {\tt CT[I*Pi*A,B] + CT[C,I*Pi*D]}}\\
\breakline\\
\outline{7}{& {\tt I*CT[Pi*A,B]}}
}

%%%%%%%%%%%%%%%%%%%%%%%%%%% 
\subsection{The Lie coalgebra of indecomposables}
\label{sec:coLie}
In applications it can be useful to focus on the most complicated part of an expression. One way to do so is to look at an expression defined modulo products. In particular, one may want to decide if two expressions are equal up to `simpler' product terms. Since MPLs form a shuffle algebra, it can be hard to decide if an expression is zero up to product terms. For example, one may be tempted to believe that the expression
$T=G(a,b;z) + G(b,a;z)$ does not involve any product terms. In this case it is easy to see that the sum in $T$ is precisely a shuffle product, so that $T$ actually vanishes modulo products. In this section we show how we can construct a map whose kernel is precisely generated by all products among MPLs.

If $H$ denotes a graded and connected Hopf algebra, then we define its \textit{space of indecomposables} $Q(H)$ as the Hopf algebra $H$ modulo all non-trivial products (i.e., products among objects of weight at least one),
\beq
Q(H) = H/(H_{>0}\cdot H_{>0})\,.
\eeq
$Q(H)$ is obviously a vector space (because it is the quotient of two vector spaces). It is, however, not an algebra (and thus not a Hopf algebra), because it was defined precisely by removing all products. 

We now construct a projector $P:H\to Q(H)$ which allows one to remove all product terms. We start by recursively defining a linear map $R$ which acts on $x\in H_{n}$, $n>0$, by~\cite{CleanPaper}
\beq\label{eq:R_map_def}
R(x) = n\,x - m(\textrm{id}\otimes R)\Delta'(x)\,,
\eeq
where $m$ is the multiplication in $H$ and $\Delta'$ is the reduced coproduct, defined below eq.~\eqref{eq:antipode_from_delta}. One can show that the kernel of $R$ is precisely generated by all non-trivial products in $H$~\cite{CleanPaper},
\beq
\textrm{Ker }R = H_{>0}\cdot H_{>0}\,.
\eeq
The projector $P=P^2$ is then obtained by correctly normalising $R$, $P(x) = \frac{1}{n}R(x)$ for $x\in H_{n}$. For example, we have
\beq\bsp\label{eq:P_01}
P(G(0,1;z)) &\,= G(0,1;z) - \frac{1}{2}\,G(0;z)\,G(1;z)\,.
\esp\eeq
We see from the previous example that the image of an MPL may contain product terms, despite the fact that the goal was to remove product terms. There is no contradiction: the elements of $Q(H)$ are equivalence classes of MPLs defined modulo product terms. The projector $P$ assigns to a function a canonical representative of its equivalence class modulo products. The product terms ensure that relations among equivalence classes are satisfied. For example, we have
\beq\bsp\label{eq:P_01}
P(G(1,0;z)) &\,= G(0,1;z) - \frac{1}{2}\,G(0;z)\,G(1;z)\\
&\, = \frac{1}{2}\,G(1,0;z) -  \frac{1}{2}\,G(0,1;z)\\
&\, = -P(G(0,1;z))\,,
\esp\eeq
In agreement with the fact that $G(1,0;z) + G(0,1;z)$ is a shuffle product, and thus vanishes modulo products.

The projector $P$ is implemented via the function {\tt ProductProjector}. For example, we can obtain eq.~\eqref{eq:P_01} from \polylogtools,
\exbox{
\inline{1}{& {\tt ProductProjector[ G[0,1,z] ]}}\\
\breakline\\
\outline{1}{& {\tt G[0,1,z] - 1/2*G[0,z]*G[1,z]}}
}

The coproduct on $H$ induces a new structure on the space of indecomposables $Q(H)$, called a \textit{Lie coalgebra}, equipped with a \textit{cobracket} $\delta: Q(H)\to Q(H)\wedge Q(H)$ defined by
\beq\label{eq:cobracket}
\delta(x) = (P\otimes P)(1-\tau)\Delta(x)\,,
\eeq
where $\tau(a\otimes b) = b\otimes a$ is the map that reverses tensors. The cobracket has recently appeared in physics in the context of scattering amplitudes in planar $\mathcal{N}=4$ SYM~\cite{Golden:2013xva,Golden:2018gtk}. We take the opportunity to make some technical comments: First, the cobracket in eq.~\eqref{eq:cobracket} is really only defined on MPLs modulo their discontinuities, i.e., we have to put to zero of powers of $\pi$ in both entries of the wedge product. Second, in applications the cobracket in eq.~\eqref{eq:cobracket} is directly defined on the MPLs themselves, i.e., on elements of $H$, while it is in principle only defined on the equivalence classes living in the space of indecomposables $Q(H)$. This is possible because the map $\delta$ in eq.~\eqref{eq:cobracket} has the property $\delta P= \delta$, i.e., the image of an element of $H$ under $\delta$ agrees with the image of its projection on $Q(H)$. For example, the cobracket of the dilogarithm is
\beq\label{eq:wedge_01}
\delta(G(0,1;z)) = -G(0;z)\wedge G(1;z)\,.
\eeq

The cobracket on MPLs is implemented via the function {\tt Cobracket}, which can be used as shown in the following example (cf.~eq.~\eqref{eq:wedge_01}),
\exbox{
\inline{2}{& {\tt Cobracket[ G[0,1,z] ]}}\\
\breakline\\
\outline{2}{& {\tt - CTW[ G[0,z],G[1,z] ]}}
}
The wedge product is implemented via the symbol {\tt CTW}.\footnote{The name of the symbol is composed of {\tt CT} for {\tt CircleTimes} and {\tt W} for {\tt Wedge}. In \texttt{StandardForm} and \texttt{TraditionalForm} this head is automatically formatted as wedge produt.} This symbol inherits its properties from the symbol {\tt CT} defining the tensor product. In particular, it is linear with respect to rational numbers and $i$, and puts to zero all occuurences of $\pi$,
\exbox{
\inline{3}{& {\tt CTW[A,2*I*G[a,z] + 1/2*G[b,z]]}}\\
\inline{4}{& {\tt CTW[A,I*Pi] + CTW[I*Pi,B]}}\\
\breakline\\
\outline{3}{& {\tt 2*I*CTW[A,G[a,z]] + 1/2*CTW[A,G[b,z]]}}\\
\outline{4}{& {\tt 0}}
}
In addition, {\tt CTW} is antisymmetric, and always reorders its arguments into a canonical order, with the correct sign,
\exbox{
\inline{5}{& {\tt CTW[B,A]}}\\
\breakline\\
\outline{5}{& {\tt - CTW[A,B]}}
}

\begin{table}
\bgfb
\multicolumn{2}{c}{\textbf{Table~\ref{tab:hopf}: The Hopf algebra of MPLs}}
\\
{\tt Delta[expr]}& Computes the coproduct (rather, the coaction) of {\tt expr}.
\\
{\tt Delta[\{i1,...,in\},expr]}& Computes the $(i_1,\ldots,i_n)$ component of the coproduct of {\tt expr}.
\\
{\tt CT[a1,\ldots,an]}& Represents the tensor $a_1\otimes\cdots\otimes a_n$.
\\
{\tt Antipode[expr]}& Computes the antipode of the symbol {\tt expr}. 
\\
{\tt ProductProjector[expr]}& Applies the projector $P$ to {\tt expr}. 
\\
{\tt Cobracket[expr]}& Computes the cobracket of {\tt expr}. 
\\
{\tt Cobracket[\{p,q\}, expr]}& Computes the component $(p,q)$ of the cobracket of {\tt expr}. 
\\
{\tt CTW[a,b]}& Represents the wedge product $a\wedge b$. 
\egfb
\textcolor{white}{\caption{\label{tab:hopf}}}
\end{table}

\section{Symbols of MPLs}
\label{sec:symbols}
% !TEX root = polylogtools.tex

\subsection{Symbols in \polylogtools}
While the coproduct on MPLs is very useful in applications, it often contains too much information. It is often useful to focus on a piece of the coproduct which is easier to work with, albeit at the price of losing some information. Such a quantity is the \textit{symbol}~\cite{ChenSymbol,Goncharov:2005sla,Brown:2009qja,Goncharov:2010jf,Duhr:2011zq,Duhr:2012fh}, which can be defined as the maximal iteration of the coproduct (modulo $i\pi$),
\beq\label{eq:symbol_def}
\cS(x) = \Delta_{1,\ldots,1}(x)\mod i\pi\,.
\eeq
The symbol contains the same information as the maximal iteration of the coproduct. It is very easy to work with, because its entries are MPLs of weight one, i.e., ordinary logarithms. For this reason it is customary to drop the log-signs when talking about the symbol, e.g.,
\beq\bsp
\Delta_{1,1}(\li{2}(z)) &\,= -\log(1-z)\otimes\log z\,,\\
\cS(\li{2}(z)) &\,= -(1-z)\otimes z\,.
\esp\eeq
Besides this notational difference, there is no difference between the symbol $\cS$ and the maximal iteration of the coproduct $\Delta_{1,\ldots,1}$.

\polylogtools\ contains a function that allows the user to turn a maximal iteration of a coproduct into a symbol (which effectively only amounts to removing the log-signs), e.g.,
\exbox{
\inline{1}{& {\tt X = Delta[\{1,1\}, PolyLog[2,z]]}}\\
\inline{2}{& {\tt ToSymbol[X]}}\\
\breakline\\
\outline{1}{& {\tt - CT[Log[1-z],Log[z]]}}\\
\outline{2}{& {\tt - CiTi[1-z, z]}}
}
There is also a function which readily combines the maximal iteration of the coproduct with the function {\tt ToSymbol}:
\exbox{
\inline{3}{& {\tt SymbolMap[PolyLog[2,z]]}}\\
\breakline\\
\outline{3}{& {\tt - CiTi[1-z, z]}}
}
Symbol tensors are represented inside the code by lists with head {\tt CiTi}.\footnote{Which is another variant of an abbreviation for {\tt CircleTimes}. In \texttt{StandardForm} and \texttt{TraditionalForm} this head is also automatically formated as tensor product.} The reason for introducing another definition for tensor products comes from the fact that the two tensors {\tt CT} and {\tt CiTi} have different linearity properties: while {\tt CT[1-z,z]} automatically evaluates to the sum {\tt CT[1,z] - CT[z,z]}, this should obviously not be the case for the symbol tensor {\tt CiTi[1-z,z]}. Instead, symbol tensors inherit their linearity properties from the logarithm function,
\begin{eqnarray}
\label{eq:sym_1}
A\otimes(\pm1)\otimes B &=& 0\,,\\
\label{eq:sym_expand}
A\otimes(x\cdot y)\otimes B &=& A\otimes x\otimes B + A\otimes y\otimes B\,.
\end{eqnarray}
Hence, whenever an entry in a list with head {\tt CiTi} is $\pm1$, it automatically evaluates to zero,
\exbox{
\inline{4}{&{\tt CiTi[A,1,B] + CiTi[C,-1,B]}}\\
\breakline\\
\outline{4}{& {\tt 0}}
}
The additivity of the symbol in eq.~\eqref{eq:sym_expand} is not applied automatically. Instead, the user can instruct \polylogtools\ to apply it whenever possible via the function {\tt SymbolExpand} 
\exbox{
\inline{5}{& {\tt S = CiTi[A,x*y,B]}}\\
\inline{6}{& {\tt SymbolExpand[ S ]}}\\
\breakline\\
\outline{5}{& {\tt CiTi[A,x*y,B]}}\\
\outline{6}{& {\tt CiTi[A,x,B] + CiTi[A,y,B]}}
}
In applications, the entries in a symbol tensor -- the so-called \textit{letters} -- are often rational functions of kinematic variables. Using the additivity in eq.~\eqref{eq:sym_expand}, one can always write such a symbol in a form where all the letters are polynomials. The function {\tt SymbolExpand} automatically maximally factors all polynomial symbol entries over the integers, so that after the application of this function all letters are irreducible polynomials over $\mathbb{Z}$. For example, we have
\exbox{
\inline{7}{& {\tt S = CiTi[1-x\^{}3,x\^{}2] }}\\
\inline{8}{& {\tt SymbolExpand[ S ]}}\\
\breakline\\
\outline{7}{& {\tt CiTi[1-x\^{}3,x\^{}2]}}\\
\outline{8}{& {\tt 2*CiTi[1-x,x] + 2*CiTi[1+x+x\^{}2,x]}}
}
Sometimes it can be useful to undo the expansion of the symbol, in order to combine expanded terms back into products or ratios. The user can instruct \polylogtools{} to attempt to combine such terms via the function {\tt SymbolFactor}
\exbox{
    \inline{9}{& {\tt SymbolFactor[ 2*CiTi[A,x,B]-CiTi[A,y,B] ]}}\\
    \breakline\\
    \outline{9}{&{\tt CiTi[A,x\^{}2/y,B]}}
}
This works best when the coefficients of the symbol tensors are integers, since factors of $1/2$ can exponentiate to unintended square roots in this procedure.
This function is particularly useful when the symbol alphabet contains factored roots of a quadratic equation that one wants to combine again. For example we can use the function {\tt SymbolFactor} to perform simplifications of the form:
\exbox{
    \inline{10}{& {\tt SymbolFactor[CiTi[A,1-Sqrt[x]]+CiTi[A,1+Sqrt[x]]}}\\
    \breakline\\
    \outline{10}{&{\tt CiTi[A,1-x]}}
}
It is possible to obtain the list of all letters in a symbol -- the \textit{symbol alphabet} -- by applying the function {\tt GetSymbolAlphabet}. For example, we have
\exbox{
\inline{11}{& {\tt S = CiTi[1-x\^{}3,x\^{}2]}}\\
\inline{12}{& {\tt GetSymbolAlphabet[ S ]}}\\
\breakline\\
\outline{11}{& {\tt CiTi[1-x\^{}3,x\^{}2]}}\\
\outline{12}{& {\tt \{x\^{}2,1-x\^{}3\}}}
}
We note here that the alphabet returned by {\tt GetSymbolAlphabet} does not necessarily consist of irreducible polynomial letters, but it simply collects all the entries in the symbol. In order to obtain an alphabet of irreducible letters, the {\tt GetSymbolAlphabet}  can be composed with the {\tt SymbolExpand} function,
\exbox{
\inline{13}{& {\tt GetSymbolAlphabet[ SymbolExpand[ S ] ]}}\\
\breakline\\
\outline{13}{& {\tt \{1-x,x,1+x+x\^{}2\}}}
}

There are various equivalent definitions of symbols in the literature. 
An important feature of the implementation of the symbol map in \polylogtools\ is that it does not put to zero symbol tensors that contain a constant letter, e.g.,
\exbox{
\inline{14}{& {\tt SymbolMap[ Log[2]*Log[x] ]}}\\
\breakline\\
\outline{14}{& {\tt CiTi[2,x] + CiTi[x,2]}}
}
Keeping constant letters in the symbol sometimes provides valuable information on the underlying function, cf.,~e.g.,~refs.~\cite{Duhr:2011zq,Duhr:2012fh}.

\polylogtools\ contains another implementation of the symbol map $\cS$ based on dissections of decorated rooted polygons attached to MPLs~\cite{Duhr:2011zq}. It can be called through the function {\tt ComputeSymbol}. The result is fully equivalent to applying the function {\tt ToSymbol} to the maximal iteration of the coproduct, e.g.,
\exbox{
\inline{15}{& {\tt ComputeSymbol[ PolyLog[2, z] ]}}\\
\inline{16}{& {\tt SymbolMap[ PolyLog[2, z] ]}}\\
\breakline\\
\outline{15}{& {\tt - CiTi[1-z,z]}}\\
\outline{16}{& {\tt - CiTi[1-z,z]}}
}
While the two implementations are fully equivalent, we mention here that (based on experience) the implementation of {\tt ComputeSymbol} is usually faster for weights less than four, whereas \texttt{SymbolMap}  works more efficiently for higher weights.\footnote{Note that \texttt{ComputeSymbol} is only implemented through weight six, while \texttt{SymbolMap} has in principle support through the same weight as \texttt{Delta}, i.e. weight twelve.}

The symbol map assigns to an MPL expression a symbol tensor. The question then naturally arises if any symbol tensor constructed from a certain alphabet can be the symbol of a function. It turns out that this is not the case in general, however, but the symbol tensor
\beq
S = \sum_{I = (i_1,\ldots,i_n)} c_I\,a_{i_1}\otimes \ldots \otimes a_{i_n}\,, \qquad c_I\in \mathbb{Q}\,,
\eeq 
is the symbol of a function, i.e., there is a function $F$ such that $\cS(F)=S$, if and only if it satisfies the \textit{integrability condition}
\beq
\sum_{I = (i_1,\ldots,i_n)} c_I\,a_{i_1}\otimes \ldots\otimes {a}_{i_{p-1}}\otimes {a}_{i_{p+2}}\otimes\ldots\otimes a_{i_n}\,d\log a_{i_p}\wedge d\log a_{i_{p+1}} = 0\,,
\eeq
for every consecutive pair of indices $(i_p,i_{p+1})$, $1\le p< n$ and $\wedge$ denotes the usual wedge product of differential forms. For every value of $p$, the integrability condition translates into a system of linear constraints on the coefficients $c_I$. \polylogtools\ can generate these constraints for a given symbol {\tt S} via the command {\tt IntegrablityCondition[ S, p ]}. We emphasise that \polylogtools\ does not attempt to solve the integrability constraints, because these linear constraints usually give rise to very large linear systems whose solution may require dedicated algorithms and/or specialised software.

\subsection{The shuffle Hopf algebra of symbols}
The symbol map is not only linear, but it also preserves the multiplication of MPLs and maps a product of MPLs to the shuffle product of their symbol tensors,
\beq
\cS(x\cdot y) = \cS(x)\shuffle\cS(y)\,.
\eeq

In Section~\ref{sec:Hopf_review} we have seen that every shuffle algebra is naturally a Hopf algebra, with the coproduct and antipode given by the deconcatenation and reversal of words, cf.~eqs.~\eqref{eq:Delta_dec} and~\eqref{eq:antipode_def}. It follows that the shuffle algebra $B$ of all symbol tensors -- integrable or not -- forms a Hopf algebra with the deconcatenation coproduct and antipode. It is then interesting to ask what is the relation between the coproduct $\Delta$ on MPLs and the deconcatenation coproduct $\Delta_{\textrm{dec}}$ on symbol tensors. The shuffle algebra $B$, however, is too large, because the image of the symbol map is the subspace of $B$ consisting of integrable symbol tensors, which we denote by $H^0B$. It is easy to check that $H^0B$ is a Hopf subalgebra of $B$. Moreover, the symbol map preserves the coproducts (and also the antipodes), i.e., $\cS$ maps the coproduct of an MPL to the deconcatenation coproduct of its symbol, and similarly for the antipode,
\beq\bsp
\Delta_{\textrm{dec}}(\cS(x))&\, = (\cS\otimes\cS)\Delta(x)\\
S_{\textrm{dec}}(\cS(x))&\, = \cS(S(x))\,.
\esp\eeq
The deconcatenation coproduct and antipode are implemented in the code via the functions {\tt DeltaDeconcatenation} and {\tt AntipodeDeconcatenation}, which act on symbol tensors as follows,
\exbox{
\inline{1}{&{\tt DeltaDeconcatenation[ CiTi[a,b,c] ]}}\\
\inline{2}{&{\tt AntipodeDeconcatenation[ CiTi[a,b,c] ]}}\\
\breakline\\
\outline{1}{&{\tt CT[1,CiTi[a,b,c]] + CT[CiTi[a],CiTi[b,c]] + }\\
&{\tt CT[CiTi[a,b],CiTi[c]] + CT[CiTi[a,b,c],1]}}\\
\outline{2}{&{\tt - CiTi[c,b,a]]}}
}

Since $H^0B$ is a graded and connected Hopf algebra, we can apply the results of Section~\ref{sec:coLie} and study its Lie coalgebra of indecomposables. First, following eq.~\eqref{eq:R_map_def} we define a linear map $\rho$ whose kernel is precisely generated by all shuffles,
\beq
\rho(a_1\otimes\ldots\otimes a_n) = n\,a_1\otimes\ldots\otimes a_n - \shuffle(\textrm{id}\otimes\rho)\Delta'_{\textrm{dec}}(a_1\otimes\ldots\otimes a_n)\,.\eeq
The previous definition can be cast in the equivalent form~\cite{Ree:1958ab,Griffing:1995ab}
\beq
\rho(a_1\otimes\ldots\otimes a_n) = \rho(a_1\otimes\ldots\otimes a_{n-1})\otimes a_n  - \rho(a_2\otimes\ldots\otimes a_n) \otimes a_1\,.
\eeq
From this map we define a projector $\Pi^2=\Pi$ by~\cite{Duhr:2011zq}
\beq
\Pi(a_1\otimes\ldots\otimes a_n) = \frac{1}{n}\,\rho(a_1\otimes\ldots\otimes a_n)\,,
\eeq 
and the cobracket on symbols is given by (cf.~eq.~\eqref{eq:cobracket})
\beq\bsp\label{eq:cobracket_symbol}
\delta_{\textrm{dec}}(a_1\otimes\ldots\otimes a_n)&\, = (\Pi\otimes\Pi)(1-\tau)\Delta_{\textrm{dec}}(a_1\otimes\ldots\otimes a_n)\\
&\, = \sum_{k=1}^{n-1}(a_1\otimes\ldots\otimes a_k)\wedge (a_{k+1}\otimes\ldots\otimes a_n)\,.
\esp\eeq
The projector $\Pi$ and the deconcatenation cobracket $\delta_{\textrm{dec}}$ are implemented via the functions {\tt ProductProjector} and {\tt CobracketDeconcatenation}. They have the same properties as their analogues acting on MPLs described in Section~\ref{sec:coLie}, so we will not describe them here in more detail. 

%%%%%%%%%%%%%%%%%%%%%%%%%%%%
\subsection{The cobracket conjecture, classical and Nielson polylogarithms}
In ref.~\cite{Goncharov:2010jf} a conjectural criterion on the symbol of a function was presented that allows one to determine if a function of weight four can be expressed in terms of classical polylogarithms only, and it was applied to the analytic result for the two-loop six-point remainder function in planar $\mathcal{N}=4$ Super Yang-Mills of ref.~\cite{DelDuca:2009au,DelDuca:2010zg}. The criterion can be concisely stated in terms of the cobracket $\delta_{\textrm{dec}}$: if $T$ is a polylogarithmic function of weight four such that its symbol satisfies  
\beq
\delta_{\textrm{dec},2,2}(\cS(T))=0\,,
\eeq
then $T$ can be expressed in terms of classical polylogarithms only. The cobracket on symbol tensors of length $n=4$ has a particularly simple form,
\beq
\delta_{\textrm{dec},2,2}(a_1\otimes a_2\otimes a_3\otimes a_4)= \frac{1}{4}\,(a_1\wedge a_2)\wedge(a_3\wedge a_4)\,.
\eeq

It is of course interesting to speculate how this criterion extends beyond weight four. Naive speculation would lead to the guess that if $T$ is a polylogarithmic expression weight $n$ such that $\delta_{\textrm{dec},p,n-p}(\cS(T))=0$ for all $1<p<n-1$, then $T$ can be expressed (possibly up to terms in the kernel of $\cS$) in terms of classical polylogarithms only. This extension, however, seems to be too naive, because already at weight five it is not known how to express the Nielsen polylogarithm $S_{2,3}(x)$ in terms of classical polylogarithms only, even though $\delta_{\textrm{dec},2,3}(\cS(S_{2,3}(x)))=0$. Here $S_{n,p}(x)$ denotes the Nielsen polylogarithm~\cite{Nielsen},
\beq
S_{n,p}(x) = (-1)^p\,G(\underbrace{0,\ldots,0}_{n\textrm{ times}},\underbrace{1,\ldots,1}_{p\textrm{ times}};x)\,.
\eeq
Instead, a folklore conjecture (which can be proven in some cases~\cite{brown_unpublished}) states that\footnote{We thank Steven Charlton for clarifying correspondance on this topic.}
\begin{quote}
If $T$ is a polylogarithmic expression of weight $n$ such that $\delta_{\textrm{dec},p,n-p}(\cS(T))=0$ for all $1<p<n-1$, then $T$ can be expressed (possibly up to terms in the kernel of $\cS$) in terms of Nielsen polylogarithms only.
\end{quote}

It would be interesting to have a sharper extension of the criterion of ref.~\cite{Goncharov:2010jf}, which allows one to determine if a function can be expressed in terms of classical polylogarithms only even at higher weight. In the following we present such a conjecture. As a starting point, we note that the cobrackets in eqs.~\eqref{eq:cobracket} and~\eqref{eq:cobracket_symbol} are related by
\beq
\delta_{\textrm{dec}}\cS = (\Pi\cS\otimes\Pi\cS)\delta\,.
\eeq
In the remainder of this section we give evidence for the following conjecture:
\begin{quote}
If $T$ is a polylogarithmic expression of weight $n$ such that $\delta_{p,n-p}(T)=0$ for all $1<p<n-1$, then $T$ can be expressed in terms of classical polylogarithms only.
\end{quote}
In the remainder of this section we present some evidence for this conjecture. In particular, we show through weight six relations that relate different Nielsen polyogarithms up to classical polylogarithms.

First we note that the conjecture is always true up to weight three, in agreement with the fact that up to weight three all MPLs can be expressed in terms of classical polylogarithms only~\cite{LewinBook,Kellerhals,Goncharov:1996}. More generally, it is known that $S_{1,n-1}(x)$ and $S_{n-1,1}(x)$ can always be expressed in terms of classical polylogarithms, and indeed we have
\beq
\delta_{p,n-p}(S_{1,n-1}(x))=\delta_{p,n-p}(S_{n-1,1}(x)) = 0\,,\textrm{  for all } 1<p<n-1\,.
\eeq

Next let us discuss the Nielsen polylogarithm $S_{2,2}(x)$. It is easy to check that $\delta_{2,2}(S_{2,2}(x))=0$. We can indeed express $S_{2,2}(x)$ in terms of classical polylogarithms only,
\beq\bsp
S_{2,2}(x) &\,= \text{Li}_4(x)-\text{Li}_4(1-x)+\text{Li}_4\left(\frac{x}{x-1}\right)-\text{Li}_3(x) \log (1-x)+\frac{1}{24} \log ^4(1-x)\\
&\,-\frac{1}{6} \log x\, \log ^3(1-x)+\frac{1}{2}\,\zeta_2\, \log ^2(1-x)+\zeta_3\, \log (1-x)+\zeta_4\,.
\esp\eeq

At weight five, we find for the first time Nielsen polylogarithms that have a non-trivial cobracket,
\beq\bsp\label{eq:S23}
\delta_{2,3}(S_{2,3}(x)) &\, = -\delta_{3,2}(S_{2,3}(x)) = \frac{1}{2}\,G(0,1;x)\wedge \zeta_3 - \frac{1}{2}\,G(1,0;x)\wedge \zeta_3\,,\\
\delta_{2,3}(S_{3,2}(x)) &\, = -\delta_{3,2}(S_{3,2}(x)) = \frac{1}{2}\,G(0,1;x)\wedge \zeta_3 - \frac{1}{2}\,G(1,0;x)\wedge \zeta_3\,.
\esp\eeq
Since classical polylogarithms of weight five lie in the kernel of $\delta_{2,3}$, we conclude from the previous equations that $S_{2,3}(x)$ and $S_{3,2}(x)$ cannot be expressed in terms of classical polylogarithms alone.
Note that, since $\cS(\zeta_3)=0$, we have,
\beq
\delta_{\textrm{dec},2,3}(\cS(S_{2,3}(x))) = \delta_{\textrm{dec},2,3}(\cS(S_{3,2}(x)))=0\,,
\eeq
and so the cobracket computed from the symbol alone does not allow one to reach this conclusion.
However, we learn something more from eq.~\eqref{eq:S23}: we see that $\delta_{2,3}(S_{2,3}(x))=\delta_{2,3}(S_{3,2}(x))$, and so the conjecture implies that the two functions are equal up to classical polylogarithms. Indeed, we find the following relation,
\beq\bsp
S_{3,2}(x) &\,= S_{2,3}(x)+\text{Li}_5(x)+\text{Li}_5(1-x)+\text{Li}_5\left(\frac{x}{x-1}\right)-\text{Li}_4(1-x)\, \log (1-x)\\
&\,+\text{Li}_4\left(\frac{x}{x-1}\right)\, \log (1-x)-\frac{1}{2}\, \text{Li}_3(x)\, \log ^2(1-x)+\frac{1}{30} \log ^5(1-x)\\
&\,-\frac{1}{8}\, \log x\, \log ^4(1-x)+\frac{1}{3}\,\zeta_2\, \log ^3(1-x)+\frac{1}{2}\, \zeta_3\, \log ^2(1-x)-\zeta_5\,.
\esp\eeq

Finally, let us analyse weight six. We find
\beq
\delta_{2,4}(S_{2,4}(x)) = \delta_{2,4}(S_{4,2}(x)) = \delta_{2,4}(S_{3,3}(x)) = 0\,,
\eeq
and
\beq\bsp
\delta_{3,3}(S_{2,4}(x)) &\, = \frac{1}{3}\,G(1,0,1;x)\wedge \zeta_3 - \frac{1}{3}\,G(1,1,0;x)\wedge \zeta_3\,,\\
\delta_{3,3}(S_{4,2}(x))&\,=- \delta_{3,3}(S_{2,4}(1-x))\,,\\
\delta_{3,3}(S_{3,3}(x))&\,=- \delta_{3,3}(S_{2,4}(x) + S_{4,2}(x))\,.
\esp\eeq
In agreement with our conjecture, we find that we can express $S_{4,2}(x)$ and $S_{3,3}(x)$ up to classical polylogarithms in terms of $S_{2,4}(x)$ and $S_{2,4}(1-x)$. The explicit expressions are
\beq\bsp
S_{4,2}(x) &\, = -S_{2,4}(1-x)-S_{2,3}(1-x)\, \log x
-\text{Li}_5(x)\, \log (1-x)
+\frac{1}{2}\, \text{Li}_4(x)\, \log ^2x\\
&\,-\frac{1}{2}\, \text{Li}_4(1-x)\, \log ^2x+\frac{1}{2}\, \text{Li}_4\left(\frac{x}{x-1}\right)\, \log ^2x
+\frac{1}{3}\, \text{Li}_3(1-x)\, \log ^3x\\
&\,-\frac{1}{12}\, \log ^3x\, \log ^3(1-x)
-\frac{1}{3}\, \zeta_3\, \log ^3x+\frac{1}{2}\, \zeta_3\, \log ^2x\, \log (1-x)
+\zeta_5\, \log (1-x)\\
&\,+2\, \zeta_5\, \log x-\zeta_2 \, \zeta_3\, \log x-\frac{1}{3}\,\zeta_2\,  \log ^3x\, \log (1-x)+\frac{1}{4}\,\zeta_2\,  \log ^2x\, \log ^2(1-x)\\
&\,+\frac{1}{2}\,\zeta_4 \, \log ^2x+\frac{1}{48}\, \log ^2x\, \log ^4(1-x)+\frac{5}{48} \,\log ^4x\, \log ^2(1-x)\\
&\,+\zeta_4 \,\log x\, \log (1-x)-\frac{1}{2}\,\zeta_3^2+\frac{3}{4}\,\zeta_6\,,
\esp\eeq
and
\beq\bsp
S_{3,3}(x) &\,= S_{2,4}(x)-S_{2,4}(1-x) +\,\text{Li}_6(1-x)-\,\text{Li}_6(x)-\,\text{Li}_6\left(\frac{x}{x-1}\right)\\
&\,-S_{2,3}(1-x)\,\log x-\text{Li}_5(1-x)\,\log(1-x)-\text{Li}_5(x)\,\log(1-x)\\
&\,-\text{Li}_5\left(\frac{x}{x-1}\right)\,\log(1-x)
+\frac{1}{2}\,\text{Li}_4(1-x)\,\log^2(1-x)\\
&\,-\frac{1}{2}\,\text{Li}_4\left(\frac{x}{x-1}\right)\,\log^2(1-x)-\frac{1}{2}\,\text{Li}_4(1-x)\,\log^2x+\frac{1}{2}\,\text{Li}_4(x)\,\log^2x\\
&\,+\frac{1}{2}\,\text{Li}_4\left(\frac{x}{x-1}\right)\,\log^2x
+\frac{1}{6}\,\text{Li}_3(x)\,\log^3(1-x)+\frac{1}{3}\,\text{Li}_3(1-x)\,\log^3x\\
&\,
-\frac{1}{72}\,\log^6(1-x)+\frac{1}{20}\,\log x\,\log^5(1-x)-\frac{1}{12}\,\log^3x\,\log^3(1-x)\\
&\,+\frac{1}{48}\,\log^2x\,\log^4(1-x)+\frac{5}{48}\,\log^4x\,\log^2(1-x)
-\frac{1}{8}\,\zeta_2\,\log^4(1-x)\\
&\,-\frac{1}{3}\,\zeta_2\,\log^3x\,\log(1-x)+\frac{1}{4}\,\zeta_2\,\log^2x\,\log^2(1-x)
-\frac{1}{6}\,\zeta_3\,\log^3(1-x)\\
&\,-\frac{1}{3}\,\zeta_3\,\log^3x+\frac{1}{2}\,\zeta_3\,\log^2x\,\log(1-x)
+\frac{1}{2}\,\zeta_4\,\log^2x+\,\zeta_4\,\log x\,\log(1-x)\\
&\,
+\zeta_5\,\log(1-x)+2\,\zeta_5\,\log x-\,\zeta_2\,\zeta_3\,\log x
-\frac{1}{2}\,\zeta_3^2-\frac{1}{4}\,\zeta_6\,.
\esp\eeq

\begin{table}
\mybgfb
\multicolumn{2}{c}{\textbf{Table~\ref{tab:symbols}: Symbols}}
\\
{\tt SymbolMap[expr]}& Computes the symbol of {\tt expr} as the maximal iteration of the coproduct.
\\
{\tt ComputeSymbol[expr]}& Computes the symbol of {\tt expr} from dissections decorated polygons.
\\
{\tt ToSymbol[expr]}& If {\tt expr} is a maximal iteration of a coproduct, then {\tt ToSymbol} turns it into a symbol expression, by removing the log-signs.
\\
{\tt CiTi[a1,\ldots,an]}& Represents the symbol tensor $a_1\otimes\ldots\otimes a_n$.
\\
{\tt SymbolExpand[sym]}& Maximally factors all the polynomial symbol letter and uses the additivity of the symbol {\tt sym} to expand them out. 
\\
{\tt SymbolFactor[sym]}& Attempts to combine terms into products and ratios in the symbol. Works best if the coefficients of the symbol are integers.
\\
{\tt GetSymbolAlphabet[sym]}& Returns the symbol alphabet of the symbol {\tt sym}. 
\\
{\tt IntegrabilityCondition[sym, p]}& Returns the constraint from the integrability condition applied to the factors $p$ and $p+1$ in the symbol {\tt sym}. 
\\
{\tt DeltaDeconcatanation[sym]}& Computes the deconcatenation coproduct of the symbol {\tt sym}. 
\\
{\tt AntipodeDeconcatanation[sym]}& Computes the deconcatenation antipode of the symbol {\tt sym}. 
\\
{\tt ProductProjector[sym]}& Applies the projector $\Pi$ to the symbol {\tt sym} and projects it to the space of indecomposables. 
\\
{\tt CobracketDeconcatenation[sym]}& Computes the cobracket attached to the deconcatenation coproduct of the symbol {\tt sym}. 
\myegfb
\textcolor{white}{\caption{\label{tab:symbols}}}
\end{table}

\section{Working with \polylogtools}
\label{sec:basic}
% !TEX root = polylogtools.tex

So far we have only discussed the implementation of the basic objects -- MPLs, their coproduct and symbols -- into \polylogtools, and we have discussed their very basic usage (e.g., how to compute the coproduct or the symbol of an MPL). In this section we describe a set of functions that can be used at runtime to manipulate expressions involving MPLs in \mathematica. Note that since \polylogtools\ automatically loads \hpl, all functions from this package are also available. For a description of these functions we refer to refs.~\cite{Maitre:2005uu,Maitre:2007kp}.

\subsection{Manipulating expressions}
\label{sec:manipulate}
We start by describing a collection of functions that are useful to manipulate MPL expressions inside \mathematica. A valid MPL expression is a polynomial built out of the functions {\tt G}, {\tt H} (or {\tt HPL}) and {\tt Li} (and the built in \mathematica\ functions {\tt PolyLog} and {\tt Log}), as well as the transcendental constants {\tt MZV}, {\tt Zeta} and {\tt Pi} (as well as certain special symbols like {\tt HPLs6} defined by \hpl~\cite{Maitre:2005uu,Maitre:2007kp}). All the functions described in this section can be applied to any valid MPL expression.

When working with big expressions, it is often useful to focus on certain subexpressions, or to read out the elementary building blocks (e.g.~MPLs) the expression is composed of. 
The function {\tt GetGs} returns a list of all {\tt G}s in the expression. Analogously, {\tt GetGArguments} returns the set of arguments of the {\tt G}, {\tt H} and {\tt HPL} objects in the given expression, while {\tt GetGIndices} returns the entries of the weight vectors of the {\tt G}, {\tt H} and {\tt HPL} objects. For example, we have
\exbox{
\inline{1}{& {\tt T = 3*G[a,z]+Pi\^{}2*G[b,x]+Zeta[3] + 2*G[a,0,x];}}\\
\inline{2}{& {\tt GetGs[ T ]}}\\
\inline{3}{& {\tt GetGArguments[ T ]}}\\
\inline{4}{& {\tt GetGIndices[ T ]}}\\
\breakline\\
\outline{2}{& {\tt \{ G[a, z], G[b, x], G[a, 0, x] \}}}\\
\outline{3}{& {\tt \{ 0, a, b, x, z \}}}\\
\outline{4}{& {\tt \{ 0, a, b \}}}
}
%{\tt TranscendentalWeight} returns the weight of a MPL or zeta value. 
The function {\tt GetWeightTerms} returns the subexpression of all terms of a given weight in an expression. For example, 
\exbox{
\inline{5}{& {\tt GetWeightTerms[ T, 1 ]}}\\
\inline{6}{& {\tt GetWeightTerms[ T, 2 ]}}\\
\inline{7}{& {\tt GetWeightTerms[ T, 3 ]}}\\
\breakline\\
\outline{5}{& {\tt 3*G[a,x]}}\\
\outline{6}{& {\tt 2*G[a,0,x]}}\\
\outline{7}{& {\tt Pi\^{}2*G[b,x]+Zeta[3]}}
}

The previous functions allow the user to focus on a specific subset of an expression. There are other functions which allow one to manipulate an expression without changing its value. In Section~\ref{sec:shuffle_stuffle} we have already encountered the functions {\tt ShuffleG}, {\tt StuffleLi}, {\tt DecomposeToLyndonWords} and {\tt ExtractZeroes}. They all fall into this category. Since they have already been discussed in Section~\ref{sec:shuffle_stuffle}, we will not discuss them here. 

Whenever $a_n\neq 0$, MPLs are invariant under a simultaneous rescaling of their arguments,
\beq\label{eq:G_scaling}
G(a_1,\ldots,a_n;z) = G(k\cdot{a_1},\ldots,k\cdot{a_n};k\cdot z)\,, \quad  a_n,k \neq 0\,.
\eeq
We can use eq.~\eqref{eq:G_scaling} to reduce all MPLs with $a_n\neq 0$ to the form $G(\vec a;1)$. If $a_n=0$, we can extract the trailing zeroes using the shuffle algebra properties before rescaling the last argument to unity. In this way we can always express any expression in terms of MPLs of the form $G(\vec a;1)$ and $G(\vec 0;z)$, $a_n,z\neq 0$. This operation is implemented in \polylogtools\ via the function {\tt NormalizeG}, e.g.,
\exbox{
\inline{8}{& {\tt NormalizeG[ T ]}}\\
\breakline\\
\outline{8}{& {\tt 3*G[a/z,1]+Pi\^{}2*G[b/x,1]+Zeta[3] + 2*(G[0,x]*G[a/x,1] - }\\
& {\tt G[0,a/x,1])}}
}

In applications the arguments of MPLs are often complicated rational or algebraic functions. It is possible to instruct \polylogtools\ to simplify the arguments of the {\tt G} functions without touching the rest of the expression. 
The utility function {\tt GArgumentSimplify} simplifies the arguments of all {\tt G}, {\tt Li}, {\tt PolyLog} and {\tt Log} objects in an expression leaving the rest of the expression unchanged.

Also the coefficients multiplying the MPLs are often functions rather than constants. It can be useful to collect the coefficients of all transcendental quantities (i.e., MPLs and MZVs). While this can in principle be done using the built-in \mathematica\ function {\tt Collect}, experience shows that this function is rather slow when acting on large expressions. The \polylogtools\ function {\tt GatherTranscendentals}, which is built around the \mathematica\ function {\tt GatherBy}, is usually much faster. It is used as shown in the following example
\exbox{
\inline{9}{& {\tt T = x*G[a,b,x] + G[a,b,x] + x*G[c,x];}}\\
\inline{10}{& {\tt GatherTranscendentals[ T ]}}\\
\breakline\\
\outline{10}{& {\tt (1+x)*G[a,b,x] + x*G[c,x]}}
}
It is possible to apply the {\tt Simplify} function to the coefficients multiplying the transcendental quantities using the function {\tt GCoefficientSimplify} (similar to {\tt Collect[..., ..., Simplify]}). It is also possible to collect an expression with respect to the non-transcendental prefactors,
\exbox{
\inline{11}{& {\tt GatherPrefactors[ T ]}}\\
\breakline\\
\outline{11}{& {\tt G[a,b,x] + x*(G[a,b,x]+G[c,x])}}
}

\begin{table}
\bgfb
\multicolumn{2}{c}{\textbf{Table~\ref{tab:basic2}: Manipulating expressions}}\\
\tt GetGs[expr] & Returns a set of all {\tt G}s in {\tt expr}.
\\
\tt GetGArguments[expr] & Returns the set of all arguments of the {\tt G}s in {\tt expr}.
\\
\tt GetGIndices[expr] & Returns the set of all entries in the weight vectors of the {\tt G}s in {\tt expr}.
\\
\tt GetWeightTerms[expr,n] & Returns the weight {\tt n} terms in {\tt expr}.
\\
\tt NormalizeG[expr] & First applies {\tt ExtractZeroes}, then replaces every $G(a_1,\ldots,a_n;z)$ in {\tt expr} with $G(\frac{a_1}{z},\ldots,\frac{a_n}{z};1)$ if $a_n,z\neq 0$.
\\
\tt GArgumentSimplify[expr] & Simplifies the arguments of all {\tt G}, {\tt Li}, {\tt PolyLog} and {\tt Log} functions in {\tt expr}.
\\
\tt GatherTranscendentals[expr] & Groups terms in {\tt expr} by transcendental object.
\\
\tt GatherPrefactors[expr] & Groups terms in {\tt expr} by rational/algebraic prefactor.
\\
\tt GCoefficientSimplify[expr] & Same as {\tt GatherTranscendentals[expr]}, but applies the {\tt Simplify} function to the coefficients.
\egfb
\textcolor{white}{\caption{\label{tab:basic2}}}
\end{table}

%%%%%%%%%%%%%%%%%%%%%%%%%%%%%%%%%%%%%%

\subsection{Series expansions of MPLs}
\label{sec:MPL_series}

Since MPLs are transcendental functions with logarithmic singularities, one is often interested in obtaining the series expansion of an MPL expression close to a (singular) point. If the function $f(z)$ has a logarithmic singularity (branch point) at the point $z=z_0$, then we cannot expand $f(z)$ into a Laurent series in any neighbourhood of the point $z_0$. Instead, $f$ admits an expansion of the form
\beq\label{eq:log_expansion}
f(z) = \sum_{k=0}^nf_k(z)\,\log^k(z-z_0)\,,
\eeq
where the functions $f_k(z)$ are analytic in a neighbourhood of $z=z_0$ and can be expanded into a Laurent series. In the case of MPLs the value of $n$ is bounded by the weight. In the following we discuss how to compute the expansion of MPLs of the form $G(\vec a;z)$ where the weight vector is independent of $z$ around the point $z=0$. In that case the expansion can be obtained in an algorithmic way which has been implemented into \polylogtools. Expansions around other points $z_0\neq 0$ can be obtained by letting $z=z_0-z'$ and working out the functional equations that map MPLs of the form $G(\vec a;z_0-z')$ to those of the form $G(\vec a';z')$. In this way the problem is reduced to finding the expansion around $z'=0$. Finding the required functional equations can often be done in an algorithmic way as well, and we refer for example to refs.~\cite{Remiddi:1999ew,Vollinga:2004sn,Maitre:2005uu,Duhr:2012fh,Duhr:2014woa} and to Section~\ref{sec:fiber}.

Let us now focus on a single MPL of the form $G(a_1,\ldots,a_n;z)$, where we assume that the $a_i$ are independent of $z$. This function has a logarithmic singularity at $z=0$ if and only if $a_n=0$. In Section~\ref{sec:shuffle_stuffle} we have seen that we can use the shuffle algebra properties to write any MPL as a linear combination of products of $G(0;z) = \log z$ and MPLs where the last element in the weight vector is non-zero. The result of this operation (implemented in \polylogtools\ via the function {\tt ExtractZeroes}) is precisely the decomposition of $G(a_1,\ldots,a_n;z)$ in eq.~\eqref{eq:log_expansion} into powers of logarithms multiplied by functions that are analytic at the origin. We have in this way reduced the problem to finding the series expansion of MPLs with $a_n\neq 0$. We can then use eqs.~\eqref{eq:Li_m_def} and~\eqref{eq:Gm_def} to obtain the expansion around $z=0$ ($a_i\neq 0$),
\beq\bsp
G&\Big(\vec{0}_{m_1-1},a_1, \ldots,\vec{0}_{m_k-1},a_k;z\Big)\\
&\, = (-1)^k\,\sum_{n=1}^\infty\frac{1}{n^{m_1}}\,\left(\frac{z}{a_1}\right)^{n}\,
Z_{m_2,\ldots,m_k}\!\left(\frac{a_1}{a_2},\ldots,\frac{a_{k-1}}{a_k};n-1\right)\,,
\esp\eeq
where the coefficients on the right-hand side are written in terms of $Z$-sums~\cite{Moch:2001zr}, which can be thought of as a variant of harmonic numbers~\cite{Vermaseren:1998uu} depending on additional variables,
\beq\label{eq:Z_sum_def}
Z_{m_1,\ldots,m_k}\!\left(x_1,\ldots,x_k;n\right) = \sum_{p=1}^n\frac{x_1^p}{p^{m_1}}\,Z_{m_2,\ldots,m_k}\!\left(x_2,\ldots,x_k;p-1\right)\,.
\eeq
The $Z$-sums are polynomials. They can be efficiently computed for every value of $n$ in \mathematica\ using the recursive definition in eq.~\eqref{eq:Z_sum_def} (and caching intermediate sums). Using these steps, we can generate series expansions of MPLs of the form $G(a_1,\ldots,a_n;z)$ up to any desired  order.

Series expansions of MPL expressions can be obtained in \polylogtools\ from the function {\tt ExpandPolyLogs} as shown in the following example,
\exbox{
\inline{1}{& {\tt ExpandPolyLogs[ G[a,0,z]/(1-z), \{z,0,2\} ]}}\\
\breakline\\
\outline{1}{& {\tt (z*(1 - G[0,z]))/a + (z\^{}2*(1 + 4*a - 2*G[0,z] - }\\
& {\tt 4*a*G[0,z]))/(4*a\^{}2) }}
 }
The argument of {\tt ExpandPolyLogs} can be any expression made out of objects for which the {\tt Series} function can obtain a series representation, as well as MPLs $G(\vec a;z)$ where the weight vector $\vec a$ is independent of $z$. \texttt{ExpandPolyLogs} automatically uses \texttt{ExtractZeroes} internally, to extract the logarithmic terms as seen in the above example.

%%%%%%%%%%%%%%%%%%%%%%%%%%%%%%%%%%%%%%

\subsection{Differentiation and integration}
\label{sec:DG_Gint}
Two of the basic operations on MPLs are differentiation and integration. In this section we discuss how these operations are implemented in \polylogtools.

Let us start by discussing derivatives. From the definition in eq.~\eqref{eq:MPL_def} it is easy to see that MPLs satisfy the differential equation
\beq
\partial_x G(a_1,\ldots,a_n;x) = \frac{1}{x-a_1}\,G(a_2,\ldots,a_n;x)\,.
\eeq
The previous equation is only true assuming that the weight vector $\vec a$ is independent of $x$. In applications one often encounters situations where one needs to compute partial derivatives with respect to the arguments of the weight vector. While it is possible to obtain closed formul\ae\ for these partial derivatives~\cite{GoncharovMixedTate}, in \polylogtools\ partial derivatives of MPLs are evaluated with the help of the coproduct. Indeed, the coproduct of a derivative $\Delta$ and the differential operator $\partial_x$ are related through the identity~\cite{Duhr:2012fh,Brown:coaction}
\beq\label{eq:Delta_der}
\Delta\partial_x = (\textrm{id}\otimes\partial_x)\Delta\,.
\eeq
Using the fact that the Hopf algebra of MPLs is graded and connected, one can obtain the following formula for the derivative of an MPL expression $F$ of uniform weight $n$,
\beq\label{eq:derivative_coproduct}
\partial_xF = m(\textrm{id}\otimes \partial_x)\Delta_{n-1,1}(F)\,.
\eeq
On the right-hand side of this equation the derivative only acts on MPLs of weight one, i.e., ordinary logarithms, for which the derivatives are easily computed. 

Partial derivatives are implemented in \polylogtools\ through the function {\tt DG}. This function takes two arguments. The first argument is the expression whose derivative one wishes to compute, and the second argument is the variable with respect to which one differentiates. Moreover, this function satisfies all the basic properties of a derivative (linearity and Leibniz rule) in its first argument, and it evaluates to the built-in \mathematica\ implementation of the derivative whenever the first argument does not depend on {\tt G}, {\tt H}, {\tt HPL} or {\tt Li} functions. When acting on an MPL, it evaluates its derivative with respect to the second argument by means of eq.~\eqref{eq:derivative_coproduct}. The following example illustrates the usage of the function {\tt DG},
\exbox{
\inline{1}{& {\tt T = 1/(1-x)*G[1/(1-x),0,0,1,1,1/(1+z)] + }\\
&{\tt (1+z)/z*G[-2,3,z];}}\\
\inline{2}{& {\tt GCoefficientSimplify[ DG[ T, x ] ]}}\\
\inline{3}{& {\tt GCoefficientSimplify[ DG[ T, z ] ]}}\\
\breakline\\
\outline{2}{&{\tt -(G[0,0,1,1,(1 + z)\^{}(-1)]/((-1 + x)*(x + z))) + }\\
& {\tt G[-(-1+x)\^{}(-1),0,1,1,(1 + z)\^{}(-1)]/(-1+x)\^{}2 + }\\
& {\tt G[-(-1+x)\^{}(-1),0,0,1,1,(1+z)\^{}(-1)]/(-1+x)\^{}2}}\\
 \outline{3}{& {\tt ((1+z)*G[3,z])/(2*z+z\^{}2) - G[-2,3,z]/z\^{}2 }\\
 & {\tt + G[0,0,1,1,(1+z)\^{}(-1)]/(x+z+x*z+z\^{}2)}}
 }

An important property of MPLs is that they behave nicely under integration. Consider the vector space generated by all functions of the form $R(z)\,G(\vec a;z)$, where the weight vector is independent of $z$ and $R(z)$ is a rational function. This vector space is closed under taking primitives, i.e., for every function $f(z)$ in that vector space there is a function $F(z)$ in the same space such that $\partial_zF=f$. The function $F$ can be found in an algorithmic way using integration by parts (see, e.g., ref.~\cite{Broedel:2017kkb}). This algorithm is implemented in \polylogtools\ through the function {\tt GIntegrate}. Its first argument is the integrand $f$ and the second argument the variable $z$. For example,  
\exbox{
\inline{4}{& {\tt T = GIntegrate[ G[0,1,z] z/(a-z)\^{}2, z ]}}\\
\inline{5}{& {\tt GCoefficientSimplify[ DG[ T , z] ]}}\\
\breakline\\
\outline{4}{& {\tt a*(-(G[0,1,z]/a) - G[0,1,z]/(-a+z) + G[a,1,z]/a) +  G[a,0,1,z]}}\\
\outline{5}{& {\tt G[0,1,z] z/(a-z)\^{}2}}
}
Whenever the weight vector depends on the integration variable $z$, or if the integrand involves non-rational algebraic functions of $z$, the algorithm does not converge and returns an unevaluated expression. We emphasise that this is not a limitation of our implementation, but in these cases the space of MPLs may not be enough to perform all the integrals and more general classes of functions, e.g. of elliptic type, may be required.

In applications one is usually not directly interested in the primitive of a function, but in its integral over a certain range. One can easily evaluate definite integrals of MPL expressions by first computing a primitive and then evaluating the primitive at the end-points of the integration range. Doing so often results in spurious singularities that cancel in the final result. For example, consider the integral
\beq
J = \int_0^b\frac{dz}{z-b}\,\left[G(b,a;z) - G(a;b)\,G(b;z) + G(a,b;b)\right]\,.
\eeq
We can formally do the integral by computing a primitive and evaluating it at $z=b$ and $z=0$,
 \beq\label{eq:I_result}
 J = G(b,b,a;b) - G(a;b)\,G(b,b;b) + G(a,b;b)\,G(b;b)\,.
 \eeq
 This last expression is ill-defined because $G(b,\ldots;b)$ is divergent. In Section~\ref{sec:shuffle_stuffle} we have seen how we can use the shuffle algebra properties to replace divergent MPLs by their shuffle regularised versions,
 \beq\bsp
   G(b;b)&\, = -\zeta_1\,,\\
  G(b,b;b)&\, = \frac{1}{2}\,\zeta_1^2\,,\\
 G(b,b,a;b)&\, = \frac{1}{2}\,\zeta_1^2\,G(a;b) - \zeta_1\,G(a;b) + G(a,b,b;b)\,.
 \esp\eeq
 It is easy to check that if the previous relations are inserted into eq.~\eqref{eq:I_result} all powers of the regulator $\zeta_1$ cancel, leaving the finite result
 \beq
 J = G(a,b,b;b)\,.
 \eeq
 All these steps are automated in \polylogtools, as shown in the following example,
 \exbox{
 \inline{6}{& {\tt J = GIntegrate[ (G[b,a,z] - G[b,z] G[a,b] + }\\
 & {\tt G[a,b,b])/(z-b), z ]  /.  z -> b}}\\
 \inline{7}{& {\tt ShuffleRegulate[ J ]}}\\
 \breakline\\
 \outline{6}{& {\tt G[b,b,a,b] - G[a,b]*G[b,b,b] + G[b,z]*G[a,b,b]}}\\
\outline{7}{& {\tt G[a,b,b,b]}}
}
In cases where the integral has not only spurious logarithmic end-point singularities but also poles, replacing MPLs by their shuffle regularised version is not sufficient, and one needs to carefully expand the MPLs around the pole. This can for example be done using the {\tt ExpandPolyLogs} function described in the Section~\ref{sec:MPL_series}.

%%%%%%%%%%%%%%%%%%%%%%%%%%%%%%%%%%%%%%%%%%%%%%%%%%%%%%%%%%

\subsection{Numerical evaluation of MPLs}
It is important to be able to evaluate MPL expressions, and for this reason a considerable effort has been put by the community in developing fast and reliable computer libraries for the numerical evaluation of (some classes of) MPLs, see for example refs.~\cite{Gehrmann:2001jv,Gehrmann:2001pz,Kalmykov:2004xg,Maitre:2005uu,Maitre:2007kp,Vollinga:2004sn,Buehler:2011ev,Frellesvig:2016ske,Frellesvig:2018lmm,Ablinger:2018sat}. 

Given the vast amount of publicly available codes for the evaluation of MPLs, \polylogtools\ does not have its own numerical routines for the evaluation of MPLs, but it relies on the \hpl\ and \ginac~\cite{Bauer:2000cp} packages. Since the \hpl\ package is loaded together with \polylogtools, {\tt HPL} functions can be evaluated numerically as described in the manual of that package~\cite{Maitre:2005uu,Maitre:2007kp}, and we will not discuss it here any further.

\ginac\ is a {\tt C++} library for numerical and symbolic computations in high-energy physics. It contains an implementation of the algorithms of ref.~\cite{Vollinga:2004sn} for the numerical evaluation of MPLs. The command-line utility {\tt ginsh} allows the user to start an interactive shell-version of \ginac. We refer to the \ginac\ manual for more details~\cite{Ginacweb}.

\polylogtools\ contains an interface that allows the user to evaluate valid MPL expressions using the interactive {\tt ginsh} environment from within \mathematica. For example,
\exbox{
\inline{1}{& {\tt T = 1/(1-x)*G[1/(1-x),0,0,1,1,1/(1+z)];}}\\
\inline{2}{& {\tt Ginsh[ T, \{x->0.3, z->0.45\} ]}}\\
\breakline\\
\outline{2}{& {\tt -0.0294179470484662503367597193416603238279}}
}
The input expression is transformed into a string that is a valid \ginac\ expression. This string is written to a temporary file that is then piped through the {\tt ginsh} shell command via the {\tt Run} function in \mathematica. The result returned by {\tt ginsh} is written to a temporary file. The content of this file is imported back into \polylogtools\ and returned by the {\tt Ginsh} function. The temporary files are then deleted from the disc. It is possible to instruct \polylogtools\ not to delete the temporary files by setting the option {\tt Debug} to {\tt True} (the default is {\tt False}) as shown in the following example:
\exbox{
\inline{3}{& {\tt Ginsh[ T, \{x->0.3, z->0.45\}, Debug -> True ]}}\\
\breakline\\
\outline{3}{& {\tt -0.0294179470484662503367597193416603238279}}
}
\ginac\ allows the user to evaluate MPLs with arbitrary precision, and the user can choose the target number of digits by setting an appropriate flag. The target precision for \ginac\ can be chosen dynamically by the user when calling the {\tt Ginsh} function by setting the option {\tt PrecisionGoal}. The default value is 30. For example, the user may increase or decrease the requested precision as shown in the following example:
\exbox{
\inline{4}{&{\tt Ginsh[ T, \{x->0.3, z->0.45\}, PrecisionGoal -> 10]}}\\
\inline{5}{&{\tt Ginsh[ T, \{x->0.3, z->0.45\}, PrecisionGoal -> 100]}}\\
\breakline\\
\outline{4}{& {\tt -0.02941794704846625}}\\
\outline{5}{& {\tt
-0.02941794704846625033675971934166032382890897}\\
&{\tt 19017828790593790231936752156076091770442309841}\\
&{\tt 11624061172247650989277119}}
}
Let us make an important comment at this point. MPLs are multi-valued functions, and so care is needed when evaluating MPLs to obtain the correct numerical value. When calling \ginac, \polylogtools\ does not make any assumption on the branch cuts of the functions and it solely relies on the choices made by \ginac. \polylogtools\ merely translates an expression into a format that can be processed by \ginac, and it is up to the user to make sure that he or she understands the branch cut structures of the functions that are evaluated before calling \ginac.

Closely related to numerical evaluation is fitting numerical constants using the PSLQ algorithm~\cite{pslq}. \polylogtools\ provides the user with two possible ways to fit transcendental constants. The function {\tt RunPSLQ} is based on the implementation of the PSLQ algorithm in \mathematica\ by P.~Bertok. The source code of this implementation is publicly available from the Wolfram \mathematica\ library~~\cite{bertok} and is shipped together with \polylogtools. In more recent versions, \mathematica\ provides a built-in routine to find integer relations via the {\tt FindIntegerNullVector} function. The \polylogtools\ function {\tt PSLQFit} relies on this implementation. 

Let us illustrate the use of these functions with an example. The transcendental constant $G(0,1;1/2) = -\li{2}(1/2)$ can be expressed as a linear combination of $\log^22$ and $\pi^2$. The coefficients of this linear combination can be found using the PSLQ algorithm from the numerical value of $G(0,1;1/2)$,
\exbox{
\inline{6}{& {\tt num = Ginsh[ G[0,1,1/2], \{\} ]}}\\
\inline{7}{& {\tt RunPSLQ[ num, \{Log[2]\^{}2, Pi\^{}2\}, 20 ]}}\\
\inline{8}{& {\tt PSLQFit[ num, \{Log[2]\^{}2, Pi\^{}2\}, 20 ]}}\\
\breakline\\
\outline{6}{& {\tt -0.582240526465012505902656320159680108746}}\\
\outline{7}{& {\tt -Pi\^{}2/12 + Log[2]\^{}2/2}}\\
\outline{8}{& {\tt -Pi\^{}2/12 + Log[2]\^{}2/2}}
}
The last argument of the {\tt RunPSLQ} and {\tt PSLQFit} functions is the number of digits that should be taken into account in the fit. It is usually advisable to evaluate the fit with several different numbers of digits, in order to ensure that the results are stable and that the identified rational coefficients are not accidental.

\subsection{Fibration bases}
\label{sec:fiber}

An important part of applying MPLs to the computation of Feynman integrals is determining a combination of MPLs whose symbol matches a given (integrable) symbol tensor. While in general this is a highly complicated task for which no algorithmic solution is known, under certain conditions on the symbol alphabet it is possible to determine an MPL expression with the given symbol in an algorithmic way. In this section we describe such a criterion, and the corresponding algorithm implemented into \polylogtools.

Consider a symbol alphabet $\mathcal{A}$. We assume that all letters are non-constant rational functions in a set of variables $x_1,\ldots, x_m$. Without loss of generality we may assume that all letters are irreducible polynomials $p_i$ over $\mathbb{Z}$. 

Next, we assume that all letters are linear in one of the variables, which we choose to be $x_1$. Then all the letters in $\mathcal{A}$ can be written in the form
\beq
p_i(x_1,\ldots,x_m) = a_i(x_2,\ldots,x_m) \, x_1+b_i(x_2,\ldots,x_m)\,.
\eeq
Following ref.~\cite{Brown:2008um}, we define the set $\mathcal{A}_{x_1}$ as the set consisting of all irreducible non-constant polynomial factors in $a_i$, $b_i$ and $a_ib_j-a_jb_i$. If all the polynomials in $\mathcal{A}_{x_1}$ are linear in one of the variables, say $x_2$, we can iterate the procedure and construct the set $\mathcal{A}_{x_1,x_2}$. We say that a symbol is \textit{linearly reducible}~\cite{Brown:2008um} if we can find an ordering of the variables (for example the natural ordering $(x_1,\ldots, x_m)$), which allows us to iterate this procedure until the end, i.e., we can find at every step a variable $x_{k+1}$ such that all polynomials in $\mathcal{A}_{x_1,\ldots,x_k}$ are linear in $x_{k+1}$. Given an integrable symbol tensor $S$ that is linearly reducible with respect to the ordering $(x_1,\ldots, x_m)$, one can find a function of the form
\beq\label{eq:canonical_integrate}
F(x_1,\ldots,x_m) = \sum_{I=(i_1,\ldots,i_m)}c_I\,G(\vec a_{1,i_1};x_1)\cdots G(\vec a_{m,i_m};x_m)\,,
\eeq
such that $\cS(F) = S$ and the weight vector $\vec a_{k,i_k}$ only involves rational functions in the variables $(x_{k+1},\ldots,x_m)$~\cite{Brown:2008um}. We call the MPLs that appear on the right-hand side of eq.~\eqref{eq:canonical_integrate} a \textit{fibration basis} for the alphabet $\mathcal{A}$ and the ordering $(x_1,\ldots x_m)$. The function $F$ can be constructed in an algorithmic way, cf.~\cite{Brown:2008um,Anastasiou:2013srw,Panzer:2014caa,Ablinger:2014yaa,Bogner:2014mha,Bogner:2015nda}. The algorithm implemented in \polylogtools\ via the function {\tt FiberSymbol} follows closely the one described in ref.~\cite{Anastasiou:2013srw}. The function {\tt FiberSymbol} takes two arguments. The first one is an integrable and linearly reducible symbol tensor, and the second one is the list of variables with the chosen ordering. This is illustrated in the following example:
\exbox{
\inline{1}{& {\tt S = CiTi[x,x-y]-CiTi[x,y]+CiTi[x-y,x]-CiTi[y,x];}}\\
\inline{2}{& {\tt FiberSymbol[ S, \{x,y\} ]}}\\
\breakline\\
\outline{2}{& {\tt G[0,y,x] + G[y,0,x]}}
}
We stress that the output of {\tt FiberSymbol} depends on the ordering of the variables. For example, if we change the ordering of the variables in the previous example, we obtain
\exbox{
\inline{3}{& {\tt FiberSymbol[ S, \{y,x\} ]}}\\
\breakline\\
\outline{3}{& {\tt -G[0,x]*G[0,y] + G[0,x]*G[x,y] + 2*G[0,0,x]}}
}
If the input symbol tensor is not integrable or not linearly reducible with respect to the ordering in the second argument, then the algorithm fails and a warning will be shown.

So far we have only described how to use fibration bases to find a function whose symbol matches a given symbol tensor. In applications it is often useful to write a given MPL expression in terms of a fibration basis. More precisely, consider an MPL expression $F$ with the property that all MPLs have a linearly reducible symbol alphabet with respect to a same ordering of the variables. Then the user can use \polylogtools\ to write $F$ in terms of the fibration basis with respect to that ordering. We illustrate this on the following example:
\exbox{
\inline{4}{& {\tt F = G[0,1,1+x,1-y];}}\\
\inline{5}{& {\tt ToFibrationBasis[ F, \{x,y\} ]}}\\
\breakline\\
\outline{5}{& {\tt Pi\^{}2*G[-1,x]/6+G[0,y]*G[0,-1,x]+G[0,-1,-1,x]-}\\
& {\tt G[-1,x]*G[1,0,y]-G[0,-1,-y,x]+G[1,0,0,y]+Zeta[3]}}
}
Unlike the output of {\tt FiberSymbol}, the output of {\tt ToFibrationBasis} involves also terms proportional to MZVs that are not detected by the symbol. These terms are reconstructed using the coproduct combined with the numerical fitting technique described in ref.~\cite{Anastasiou:2013srw}, where the MZVs are reconstructed by evaluating constant combinations of MPLs numerically at a single point. So far this fitting technique has been implemented through weight six, and therefore the use of {\tt ToFibrationBasis} is limited to expression of weight up to five. Since MPLs are multi-valued functions, the result of the fit can depend crucially on the numerical value used in the fit. By default, each variable is assigned a random value in the range $[0,1]$. The user can change the numerical values used in the fit through the option {\tt FitValue}:
\exbox{
\inline{6}{& {\tt ToFibrationBasis[ F, \{x,y\}, }\\
& \qquad {\tt FitValue -> \{ x-> 0.1, y->0.2\} ]}}\\
\breakline\\
\outline{6}{& {\tt Pi\^{}2*G[-1,x]/6+G[0,y]*G[0,-1,x]+G[0,-1,-1,x]-}\\
& {\tt G[-1,x]*G[1,0,y]-G[0,-1,-y,x]+G[1,0,0,y]+Zeta[3]}}
}
It is strongly recommended that the user always chooses the numerical value in a way which reflects the applications which he or she has in mind (e.g., in applications the variables are often related to kinematic variables, which take values in a certain range). This is particularly important when the symbol contains letters such as $x-y$ that could lead to polylogarithms that can develop imaginary parts depending on the relative values used for fitting the constants. For large expressions, the reduction to a fibration basis can take a considerable amount of time. It it possible to save intermediate results to the disc by setting the option {\tt Save -> "file.m"}. Intermediate results can later be read in via the option {\tt Input -> "file.m"}.

The function {\tt ToFibrationBasis} is one of the most important features of \polylogtools. In particular, fibration bases play an important role in the computation of Feynman parameter integrals via direct integration, using ideas similar to those described in refs.~\cite{Brown:2008um,Anastasiou:2013srw,Panzer:2014caa,Ablinger:2014yaa,Bogner:2014mha,Bogner:2015nda}. Indeed, if the set of polynomials in the denominator of an integrand is linearly reducible with respect to a given ordering, then we can perform all the integrations one-by-one in that order. At each step one can use the function {\tt ToFibrationBasis} to write all MPLs in the integrand in a form where the integration variable appears in the last argument. Once that is achieved, the integral can be performed easily in an algorithmic way using the {\tt GIntegrate} function. 

Another important application of fibration bases is the derivation of transformation formul\ae\ for MPLs. As an example, consider the function $G(-1,-y;x)$, and imagine we want to obtain its series expansion around $y=1/2$ with $0<x<1$. In Section~\ref{sec:MPL_series} we have seen how to use the function {\tt ExpandPolyLogs} to expand MPLs of the form $G(\vec a;z)$ around $z=0$. Hence, if we let $y=1/2-z$ and we pass to a fibration basis with respect to the ordering $\{z,x\}$, we can easily obtain the desired expansion. The corresponding piece of code is shown below:
\exbox{
\inline{7}{& {\tt F = G[-1,-y,x];}}\\
\inline{8}{& {\tt Ffiber = ToFibrationBasis[ F/\!.\,y->1/2-z, \{z,x\}, }\\
& {\tt \qquad FitValue -> \{x->0.2,z->0.01\}];}}\\
\inline{9}{& {\tt ExpandPolyLogs[ Ffiber, \{z,0,0\} ]}}\\
\breakline\\
\outline{9}{& {\tt G[-1,-1/2,x] + z*(4*G[-1,x]-2*G[-1/2,x])+}\\
& {\tt z\^{}2*(-2+2/(1+2*x)+2*G[-1/2,x])}}
}

Let us conclude by mentioning that the criterion of linear reducibility described above is only a sufficient, but not necessary, condition to find a function that matches a given symbol tensor. The sets $\mathcal{A}_{x_1,\ldots,x_k}$ provide an upper bound on the singularities that can appear after the variables $x_1,\ldots,x_k$ have been integrated out. There are more refined criteria that allow one to obtain a more refined bound on the singularities, and therefore to find fibration bases for larger classes of functions~\cite{Brown:2008um,Brown:2009ta}. These more refined criteria, however, have not yet been implemented into \polylogtools.

\begin{table}
\bgfb
\multicolumn{2}{c}{\textbf{Table~\ref{tab:basic3}: Manipulating expressions (2)}}\\
\tt ExpandPolyLogs[expr,\{x,0,n\}] & If {\tt expr} involves only MPLs of the form $G(\vec a;x)$, with $\vec a$ independent of $x$, then {\tt expr} is expanded into a series around $x=0$ up to order $n$.
\\
\tt Ginsh[expr, list] & Uses \ginac\ to evaluate {\tt expr} numerically. {\tt list} is a replacement list that specifies the numerical values that should be assigned to all the variables in {\tt expr}. {\tt Ginsh} has an option {\tt Debug}. If set to {\tt True} (which is the default), the temporary files containing the \ginac\ code are not deleted from the disc.
\\
\tt RunPSLQ[num, list, prec] & Uses the PSLQ implementation by P. Bertok with precision {\tt prec} to express {\tt num} as a rational linear combination of the quantities in {\tt list}. The argument {\tt num} must be real.
\\
\tt PSLQFit[num, list, prec] & Uses the {\tt FindIntegerNullVector} function to express {\tt num} as a rational linear combination of the quantities in {\tt list}. The argument {\tt num} can be either real or complex. The integer {\tt prec} specifies that all floats should be interpreted as having a precision of {\tt prec} digits.
\\
\tt DG[expr, x] & Computes the partial derivative of {\tt expr} with respect to {\tt x}.
\\
\tt GIntegrate[expr, x] & Computes the primitive of {\tt expr} with respect to {\tt x}. Only expressions that contain rational functions of {\tt x} and MPLs of the form $G(\vec a;x)$, with $\vec a$ independent of $x$ are allowed inside {\tt expr}.
\\
\tt ShuffleRegulate[expr] & Replaces all divergent MPLs in {\tt expr} by their shuffle-regularised version.
\egfb
\textcolor{white}{\caption{\label{tab:basic3}}}
\end{table}

\begin{table}
\bgfb
\multicolumn{2}{c}{\textbf{Table~\ref{tab:basic4}: Manipulating expressions (3)}}\\
\tt FiberSymbol[sym, list] & Returns a combination of MPLs in a fibration basis with respect to the variables and the ordering specified in list whose symbol matches {\tt sym}. The symbol {\tt sym} is assumed integrable and linearly reducible.
\\
\tt ToFibrationBasis[expr, list] & Returns {\tt expr} in a fibration basis with respect to the variables and the ordering specified in list. This function has several options decribed below.
\\
\tt FitValue & Option of {\tt ToFibrationBasis}. A list of replacements which specify the numerical values of the all the variables used in the numerical fit. The default is {\tt Automatic}, which assigns random values between 0 and 1 to each variable.
\\
\tt Save & Option of {\tt ToFibrationBasis}. The value is a string (default: the empty string). If the string is non-empty, then intermediate results of {\tt ToFibrationBasis} are written to the file whose name is the specified string.
\\
\tt Input & Option of {\tt ToFibrationBasis}. The value is a string (default: the empty string). If the string is non-empty, then the file specified by the string is read in.
\\
\tt ProgressIndicator & Option of {\tt ToFibrationBasis}. If set to {\tt True}, a dynamically updated text indicates how many MPLs still need to be converted to the fibration basis.
\egfb
\textcolor{white}{\caption{\label{tab:basic4}}}
\end{table}

\section{Single-valued MPLs}
\label{sec:sv_mpls}
% !TEX root = polylogtools.tex

Just like the logarithm function, MPLs are multi-valued functions. It is possible to define a variant of MPLs that are real-analytic and single-valued, while preserving most of their algebraic properties. The price to pay is that these functions are no longer holomorphic, i.e., they depend explicitly on the complex conjugated variables. More precisely, \emph{single-valued MPLs} are combinations of MPLs and their complex conjugates such that all discontinuities cancel. The simplest example of such a function is the single-valued version of the logarithm, which is simply given by the logarithm of the absolute value of its argument,
\beq
\log|z|^2 = \log z + \log \bar{z}\,.
\eeq
It is possible to determine the combinations that lead to single-valued MPLs in an algorithmic way. This was done for the first time in ref.~\cite{BrownSVHPLs}, where the single-valued versions of HPLs whose weight vectors only contain 0's and 1's have been defined by constructing single-valued solutions to a unipotent differential equation. This construction was extended in ref.~\cite{BrownSVMPLs} to general classes of hyperlogarithms. In refs.~\cite{Brown:2013gia,Brown:coaction} it was shown how to define general single-valued MPLs and also MZVs.

Single-valued MPLs are not just of interest in pure mathematics, but they also appear in loop computations. In particular, they show up in the computation of Feynman integrals with massless propagators and three off-shell external legs~\cite{Chavez:2012kn}, as well as in the computation of conformal four-point functions in four dimensions~\cite{Drummond:2012bg,Drummond:2013nda,Schnetz:2013hqa,hyperlog,Schnetz:2016fhy}. In addition, single-valued versions of MPLs are known to describe the multi-Regge limit of scattering amplitudes in planar $\mathcal{N}=4$ Super Yang Mills~\cite{Dixon:2012yy,DelDuca:2016lad} and the high-energy limit of the dijet cross section in QCD~\cite{DelDuca:2013lma,DelDuca:2017peo}. They also appear in the analytic result for the three-loop corrections to the soft anomalous dimension~\cite{Almelid:2015jia,Almelid:2017qju}. 

In the remainder of this section we present the implementation of single-valued MPLs in \polylogtools. 

\subsection{Single-valued MPLs in \polylogtools}

In refs.~\cite{BrownSVHPLs,BrownSVMPLs,Brown:2013gia,Brown:coaction} (see also ref.~\cite{DelDuca:2016lad}) a map ${\bf s}$ was constructed which associates to $G(\vec a;z)$ its single-valued version $\cG(\vec a;z)$. This map can be given explicitly in terms of the coproduct and the antipode on MPLs (see Section~\ref{sec:coproduct}),
\beq\label{eq:sv_map}
{\bf s}(x) = m(\tilde{S}\otimes\textrm{id})\Delta(x)\,,
\eeq
where $\tilde{S}$ is related to the the antipode of the complex conjugate of $x$, up to a sign,
\beq
\tilde{S}(x) = (-1)^{|x|}\,S(\bar{x})\,,
\eeq
and $|x|$ is the weight of $x$. The map ${\bf s}$ is obviously linear, and it also preserves the multiplication,
\beq\label{eq:sv_mult}
{\bf s}(x_1\cdot x_2) = {\bf s}(x_1)\cdot {\bf s}(x_2)\,.
\eeq
The image of any MPL expression under {\bf s} is single-valued. Note that ${\bf s}$ does not only act on functions, but also on numbers. In particular, it sends to zero all powers of $\pi$, and acts on odd MZVs of depth one in a simple way,
\beq
{\bf s}(\pi) = 0 \textrm{~~~and~~~}{\bf s}(\zeta_{2n+1}) = 2\zeta_{2n+1}\,.
\eeq

The single-valued MPLs $\cG(a_1,\ldots,a_n;z)$ are represented in \polylogtools\ by the symbols {\tt cG[a1,\ldots,an,z]}.
The map ${\bf s}$ is implemented\ as the function {\tt SV}, which can act on any MPL expression and returns its single-valued version. We illustrate this with the following example,
\exbox{
\inline{1}{& {\tt T = G[0,0,1,z] + 4*Pi\^{}2*G[1,z] + 3*Zeta[3];}}\\
\inline{2}{& {\tt SVT = SV[ T ]}}\\
\breakline\\
\outline{2}{& {\tt cG[0,0,1,z] + 6*Zeta[3]}}
}
Via eq.~\eqref{eq:sv_map}, single-valued MPLs can be expressed as linear combinations of products of MPLs and their complex conjugates. It is possible to replace all {\tt cG} objects by these combinations as shown in the following  example,
\exbox{
\inline{3}{& {\tt cGToG[ SVT ]}}\\
\breakline\\
\outline{3}{& {\tt G[0,0,1,z]+G[1,0,0,Conjugate[z]]+G[1,Conjugate[z]]*G[0,0,z]+}\\
& {\tt G[0,z]*G[1,0,Conjugate[z]]+6*Zeta[3]}}
}

It is possible to work with the {\tt cG} functions in very much the same way in \polylogtools\ as with ordinary MPLs. All functions to manipulate MPL expressions described in Sections~\ref{sec:shuffle_stuffle} and~\ref{sec:manipulate} can be used in the same way with {\tt G} and {\tt cG} functions. In particular, since the map {\bf s} respects the multiplication (cf.~eq.~\eqref{eq:sv_mult}) single-valued MPLs form a shuffle algebra, and we can decompose them into a Lyndon word basis, extract trailing zeroes, etc., just like for the {\tt G} functions.

It is also possible to differentiate and integrate single-valued MPLs. The map {\bf s} commutes with holomorphic differentiation~\cite{BrownSVHPLs,BrownSVMPLs,Brown:coaction}, $\partial_z{\bf s} = {\bf s}\partial_z$, so that we can compute derivatives in the same way as for ordinary MPLs. In particular, the function {\tt DG} can be used just like for {\tt G} functions to compute the holomorphic derivatives of single-valued MPLs. It is also possible to evaluate single-valued MPLs by means of the {\tt Ginsh} function, for example,
\exbox{
\inline{4}{& {\tt Ginsh[ SVT, \{z -> 0.1 + 0.2*I\} ]}}\\
\breakline\\
\outline{4}{& {\tt 6.382671043967572457430846014836318794 + }\\
& {\tt 1.60916123633383464361805920171367279381*I}}
}

While the map {\bf s} commutes with holomorphic differentiation, the situation is slightly more subtle when it comes to integration. Indeed, let us compute the following (holomorphic) primitive, and use the fact that single-valued MPLs can be written as ordinary MPLs for which we know how to compute primitives. We find
\begin{align}
\nonumber\int\frac{dz}{z-1}\,\cG(0;z) &\,= \int\frac{dz}{z-1}\,\left[G(0;z) + G(0;\bar{z})\right] \\
&\,= G(1,0;z) + G(0;\bar{z})\,G(1;z) \\
\nonumber&\,\neq \cG(1,0;z) = G(1,0;z) + G(0;\bar{z})\,G(1;z) + G(0,1;\bar{z})\,.
\end{align}
We see that the primitive of a single-valued function, if computed in a naive way, is not single-valued, and the difference is precisely the antiholomorphic function $G(0,1;\bar{z})$. We should keep in mind, however, that the primitive of a function is not unique. In particular, we can add to a holomorphic primitive any antiholomorphic function, and so we can find a single-valued primitive, albeit not the one that we would have naively constructed. This is a general fact: similarly to ordinary MPLs, the space of single-valued MPLs (multiplied by rational functions) is closed under taking primitives. The single-valued primitive of a single-valued MPL expression can be computed in \polylogtools\ via the {\tt cGIntegrate} function. This function works in the same way as its non-single-valued analogue {\tt GIntegrate} described in Section~\ref{sec:DG_Gint}, and so we do not describe it here any further.

In Section~\ref{sec:hopf_mpl} we have seen that ordinary MPLs modulo their discontinuities form a Hopf algebra. Since we have a map that assigns to an MPL its single-valued version, it is natural to ask if the single-valued MPLs themselves form a Hopf algebra. This is indeed the case, and single-valued MPLs form a graded and connected Hopf algebra with a coproduct $\Delta^{\textrm{sv}}$ and antipode $S^{\textrm{sv}}$ given by the same formulas as the coproduct $\Delta$ and the antipode $S$ on MPLs of Section~\ref{sec:hopf_mpl}, except that all MPLs need to be replaced by their single-valued versions. The coproduct $\Delta^{\textrm{sv}}$ and the antipode $S^{\textrm{sv}}$ are implemented in \polylogtools\ via the functions {\tt DeltaSV} and {\tt AntipodeSV}. These functions are completely analogous to the functions {\tt Delta} and {\tt Antipode} of Section~\ref{sec:hopf_mpl}.

Since both coproducts $\Delta$ and $\Delta^{\textrm{sv}}$ are computed via very similar formulas, we find it important to quickly compare the two coproducts. We do this on the example of the function 
\beq\label{eq:cG_Delta_example}
\cG(0,1;z) = G(0,1;z) + G(1;\bar{z})\,G(0;z) + G(1,0;\bar{z})\,.
\eeq
The coproduct $\Delta^{\textrm{sv}}$ acts on single-valued MPLs through the same formula as $\Delta$ on ordinary MPLs. We the find
\beq
\Delta^{\textrm{sv}}(\cG(0,1;z)) = \cG(0,1;z)\otimes 1 + 1\otimes \cG(0,1;z) + \cG(1;z)\otimes \cG(0;z)\,.
\eeq
In particular, by construction, $\Delta^{\textrm{sv}}$ only involves single-valued MPLs. The coproduct $\Delta$, instead, acts on the MPLs on the right-hand side of eq.~\eqref{eq:cG_Delta_example}. We find
\beq\bsp
\Delta(\cG(0,1;z)) &\,= \cG(0,1;z)\otimes 1 + 1\otimes \cG(0,1;z) + \cG(1;z)\otimes G(0;z) \\
&\,+ \cG(0;z)\otimes G(1;\bar{z})\,.
\esp\eeq
We see that for $\Delta$ only the first entries in the coproduct are single-valued. The two coproducts are thus genuinely different.

Let us now discuss single-valued fibration bases. Since the map {\bf s} preserves the algebra structure, single-valued MPLs satisfy  the same relations as their non-single-valued analogues (with all factors of $\pi$ removed). This implies in particular that single-valued MPLs can be expressed in terms of a fibration basis under the same conditions and in the same way as ordinary MPLs, simply by acting with {\bf s} on the corresponding relation among ordinary MPLs. For this reason, the function {\tt ToFibrationBasis} can be applied in the same way to single-valued MPLs as to ordinary ones as described in Section~\ref{sec:fiber}.

We conclude by mentioning that more general classes of single-valued MPLs show up in Feynman integral computations, where the symbol letters are neither holomorphic nor anti-holomorphic~\cite{Chavez:2012kn,Drummond:2013nda,Schnetz:2013hqa,Schnetz:2016fhy,DelDuca:2017peo}. While it is also possible to use \polylogtools\ to construct these functions using the algorithm of ref.~\cite{Schnetz:2016fhy} (cf.,~e.g.,~ref.~\cite{Chavez:2012kn,Drummond:2013nda,DelDuca:2017peo}), this algorithm is not implemented in \polylogtools\ in an automated way. We mention, however, that these functions can be obtained in an automated way from the Maple package {\sc Hyperlog Procedures}~\cite{hyperlog}.

%%%%%%%%%%%%%%%%%%%%%%%%%%%%%%%%%%%%%%%%%%%%%

\subsection{The Lie coalgebra of clean single-valued functions}
\label{sec:clean_sv}

Since single-valued MPLs form a graded and connected Hopf algebra, we can use the results of Section~\ref{sec:coLie} and construct a projector $P^{\textrm{sv}}$ to the Lie coalgebra of indecomposables of this Hopf algebra. The projector is constructed recursively following eq.~\eqref{eq:R_map_def} using the coproduct. We first define the map
\beq
R^{\textrm{sv}}(x) = n\,x - m(\textrm{id}\otimes R^{\textrm{sv}})\Delta^{\textrm{sv}\prime}(x)\,,
\eeq
where $x$ has weight $n$ and we let $P^{\textrm{sv}}(x) = \frac{1}{n}\,R^{\textrm{sv}}$.

In ref.~\cite{CleanPaper} the following \textrm{clean} single-valued MPLs have been defined,
\beq
\cC(a_1,\ldots,a_n;z) = P^{\textrm{sv}}(\cG(a_1,\ldots,a_n;z))\,.
\eeq
The images of single-valued MPLs under the projector $P^{\textrm{sv}}$ have very interesting properties. In particular, they satisfy \textrm{clean} functional relations, i.e. the same relations as the $\cG$ functions, but with all product terms removed~\cite{CleanPaper}. The clean single-valued functions are represented by the symbols {\tt cC[a1,\ldots,an,z]}, and the projector $P^{\textrm{sv}}$ is called {\tt ProductProjectorSV}. The clean single-valued functions can be expressed in terms of single-valued MPLs. This can be achieved via the function {\tt cCTocG}. Moreover, all the functions from Sections~\ref{sec:shuffle_stuffle}, \ref{sec:manipulate} and~\ref{sec:fiber} can also be applied to the clean versions. Finally, there is a cobracket $\delta^{\textrm{sv}} = (P^{\textrm{sv}}\otimes P^{\textrm{sv}})(1-\tau)\Delta^{\textrm{sv}}$ acting on these functions. The cobracket is implemented via the function {\tt CobracketSV}.

\begin{table}
\bgfb
\multicolumn{2}{c}{\textbf{Table~\ref{tab:basic4}: Manipulating expressions (3)}}\\
\tt cG[a1,\ldots,an,z] & The single-valued MPL $\cG(a_1,\ldots,a_n;z)$.
\\
\tt cC[a1,\ldots,an,z] & The clean single-valued MPL $\cC(a_1,\ldots,a_n;z)$.
\\
\tt SV[expr] & Applies the map {\bf s} to {\tt expr}.
\\
\tt cGToG[expr] & Replaces all {\tt cG} functions in {\tt expr} by ordinary MPLs and their complex conjugates.
\\
\tt cCTocG[expr] & Replaces all {\tt cC} functions in {\tt expr} by single-valued MPLs.
\\
\tt cCToG[expr] & Equivalent to {\tt cGToG[cCTocG[expr]]}.
\\
\tt cGIntegrate[expr, x] & Computes the single-valued primitive of {\tt expr} with respect to {\tt x}. Only expressions that contain rational functions {\tt x} and MPLs of the form $\cG(\vec a;x)$, with $\vec a$ independent of $x$ are allowed inside {\tt expr}.
\\
\tt DeltaSV[expr] & Applies the coproduct $\Delta^{\textrm{sv}}$ to {\tt expr}.
\\
\tt AnitpodeSV[expr] & Applies the antipode $S^{\textrm{sv}}$ to {\tt expr}.
\\
\tt ProductProjectorSV[expr] & Applies the projector $P^{\textrm{sv}}$ to {\tt expr}.
\\
\tt CobracketSV[expr] & Applies the cobracket $\delta^{\textrm{sv}}$ to {\tt expr}.
\egfb
\textcolor{white}{\caption{\label{tab:basic4}}}
\end{table}

\section{Validation}
\label{sec:validation}
% !TEX root = polylogtools.tex

\polylogtools\ has already been applied to many computations involving MPLs that have led to publications in peer-reviewed journals, both in physics and in mathematics. In this section we review these applications. The reason for doing this is twofold. First, these applications show the versatility of the code, and they can serve as examples to the variety of problems to which \polylogtools\ can be applied. Second, these examples serve at the same time as validation of the code.

\polylogtools\ was applied for the first time in ref.~\cite{Duhr:2012fh}, where the coproduct on MPLs was used to simplify the two-loop amplitudes for the production of a Higgs boson in association with three partons~\cite{Gehrmann:2011aa}.

The most prominent application of \polylogtools\ is the computation of the N$^3$LO corrections to Higgs production in gluon-fusion~\cite{Anastasiou:2014vaa,Anastasiou:2015ema}, where the package was used extensively to compute boundary conditions for differential equations~\cite{Anastasiou:2013srw,Anastasiou:2015yha} and to evaluate phase space integrals for single-emission contributions~\cite{Anastasiou:2013mca,Duhr:2014nda}. It was also used in ref.~\cite{Buehler:2013fha} to evaluate convolution integrals that appear in the mass-factorisation formulas beyond LO. Further phenomenological applications to Higgs physics include the computation of the two-loop amplitudes for Higgs production in the Standard Model Effective Field Theory~\cite{Deutschmann:2017qum} and the top-Yukawa contributions to $bbH$ production~\cite{Deutschmann:2018avk}. 

\polylogtools\ was successfully used to compute integrated counterterms for the {\sc CoLoRFulNNLO} subtraction method~\cite{DelDuca:2015zqa,DelDuca:2016csb,DelDuca:2016ily}. It was also used in the context of Lattice QCD in the computation of certain one-loop integrals that appear in the massive momentum-subtraction scheme~\cite{Boyle:2016wis}. More recently, the package was used to manipulate MPLs that appear in the computation of four and five-gluon scattering using the numerical unitarity approach~\cite{Abreu:2017xsl,Abreu:2017hqn,Abreu:2018jgq,Abreu:2018zmy}.

The package was not only applied to higher-order computations relevant to collider physics, but it has also seen various more formal applications. In particular, it has played a crucial role in the analytic computation of the three-loop soft anomalous dimension matrix~\cite{Almelid:2015jia}. It has also been used to compute certain two-loop triangles~\cite{Chavez:2012kn} and conformal four-point functions~\cite{Drummond:2013nda} in terms of single-valued MPLs. Single-valued MPLs are known to show up in multi-Regge kinematics, and \polylogtools\ was also successfully applied in that context, both in planar $\mathcal{N}=4$ SYM~\cite{Dixon:2012yy,DelDuca:2016lad,DelDuca:2018hrv,Marzucca:2018ydt} and in QCD~\cite{DelDuca:2013lma,DelDuca:2017peo}. Recently, it was used to manipulate analytic expressions for two-loop form factors in $\mathcal{N}=4$ SYM~\cite{Brandhuber:2018xzk,Brandhuber:2018kqb}. The package was also used to compute and analyse the cuts of one and two-loop integrals~\cite{Abreu:2014cla,Abreu:2015zaa,Abreu:2017ptx} and to study their relationship to the coaction of Feynman integrals~\cite{Abreu:2017enx,Abreu:2017mtm}. Finally, (a private extension of) \polylogtools\ was used to evaluate various Feynman integrals that evaluate to elliptic generalisations of MPLs~\cite{Broedel:2017siw,Broedel:2018iwv,Broedel:2018qkq,Broedel:2019hyg}.

\polylogtools\ was not only used to perform research in theoretical and mathematical physics, but it has led to new results in pure mathematics. It was applied in refs.~\cite{CharltonThesis,CharltonDan} to study reduction identities for MPLs to lower depth, and in refs.~\cite{Gangl1,Gangl2} to study some properties of MPLs of weight four, in particular to derive a novel functional relation for MPLs of weight four involving more than 100 terms. Finally, it was used in ref.~\cite{CleanPaper} to define and study some of the clean single-valued MPLs reviewed in Section~\ref{sec:clean_sv}.

\section{An example calculation}
\label{sec:example}
% !TEX root = polylogtools.tex

\begin{figure}[ht!]
    \label{fig:box}
    \center
    \includegraphics[width=0.3\textwidth]{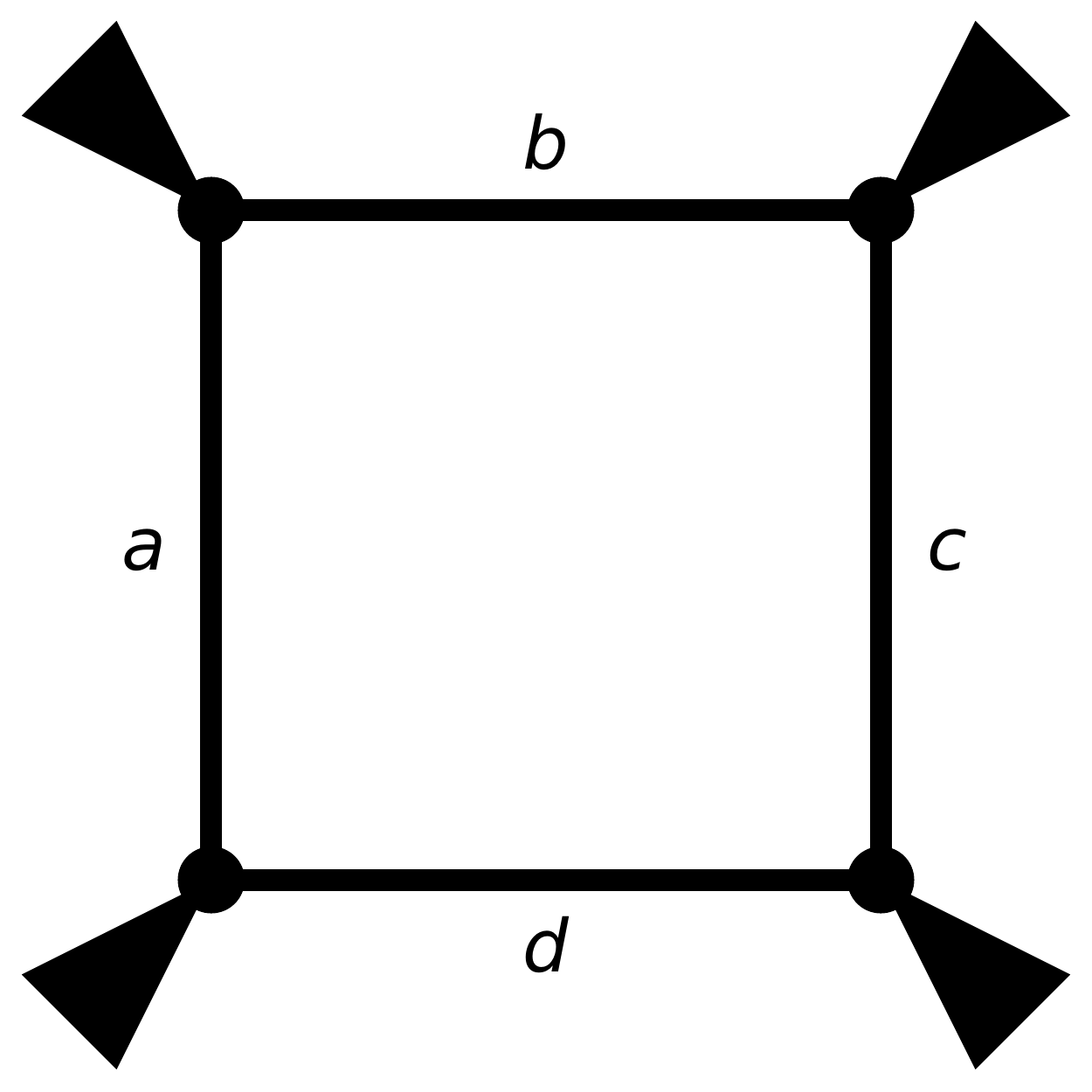}
    \caption{The four-mass box integral.}
\end{figure}

In the following we present a sample computation using \plt{} that illustrates the most important features of the package. We compute the one-loop four-mass scalar box integral shown in fig.~\ref{fig:box}. This integral is finite and furthermore one of the simplest and most prominent examples of dual-conformal integrals appearing for example in $\mathcal{N}=4$ SYM. We stress that these properties are by no means prerequisites for the use of \plt{}, in particular the package has been used to compute many infrared-divergent and non-dual-conformal integrals as pointed out in the previous section. However, this integral allows us to focus on the features of \plt{} and elaborate on some potential obstructions in the calculation and how to avoid them using \plt{}.
This example is also described in the notebook \texttt{integration\_manual.nb} distributed with the source code of the package.

The four-mass box integral can be written as
\begin{equation}
I^{4m}(u,v) = \int d^4\ell \frac{(a,c)(b,d)\Delta_6[u,v]}{(\ell,a)(\ell,b)(\ell,c)(\ell,d)}\,.
\end{equation}
Here we have used the notation $(a,b) = (x_b-x_a)^2$ that is inspired by the embedding space formalism. The four dual points $a,b,c,d$ are non-light-like separated, so that we can define the two finite dual-conformal cross ratios
\begin{equation}
u = \frac{(a,b)(c,d)}{(a,c)(b,d)}\,\quad\textrm{and}\quad v = \frac{(a,d)(b,c)}{(a,c)(b,d)}\,,
\end{equation}
as well as the Gram determinant
\begin{equation}
\Delta_6 = \sqrt{(1-u-v)^2-4uv}\,.
\end{equation}
While it is possible to linearize the Gram determinant by introducing new variables, we will refrain from doing so here, in order to illustrate how to handle the appearance of square roots in the calculation.
The starting point for our calculation is the Feynman parametrization of the
loop integral, that we write as~\cite{Bourjaily:2019jrk},
\exbox{
\inline{1}{& {\tt f1 = a[1]*u + a[2] + a[3]*v;}}\\
\inline{2}{& {\tt f2 = a[1]*a[2] + a[1]*a[3] + a[2]*a[3]; }}\\
\inline{3}{& {\tt F = d6/2/f1/f2; }}
}
Here the variable \texttt{d6} denotes the Gram determinant $\Delta_6$.
The integral over the Feynman parameters $a_i$ is over $\mathbb{P}^3$, and we can choose a chart by setting any one of the $a_i$ to one and integrating the remaining ones over the range $[0,\infty)$. We can then compute the primitive in the first variable as
\exbox{
\inline{4}{& {\tt P = GIntegrate[F/.a[3]->1,a[1]]; }}
}
which yields
\begin{equation}
P = -\frac{\Delta_6\Big[G\left(-\frac{\alpha_2+v}{u};\alpha_1\right)-G\left(-\frac{\alpha_2}{\alpha_2+1};\alpha_1\right)\Big]}{2 \left(\alpha_2^2+\alpha_2-\alpha_2 u+\alpha_2 v+v\right)}\,.
\end{equation}
Since none of the resulting polylogarithms have a $0$ in their last index, we can immediately set $\alpha_1$ to zero to see that the integral vanishes on the lower boundary.
To derive the result at the upper bound, we need to map $+\infty$ to zero, which we can do by letting $\alpha_1\to1/t$. Now we can instruct \plt{} to derive the functional identities that are required to express the primitive as a function of $t$ using
\exbox{
    \inline{5}{& {\tt Pp = ToFibrationBasis[P/.a[1]->1/t,\{t,a[2],u,v\}];}}}
which yields
\begin{equation}\begin{split}
    \frac{\Delta_6}{2\left(\alpha_2^2+(1-u+v)\alpha_2+v\right)}\Big[G\left(-1;\alpha_2\right)-G\left(0;\alpha_2\right)+G\left(-v;\alpha_2\right)\\
    -G\left(0;u\right)+G\left(0;v\right)+G\left(-\tfrac{1+\alpha_2}{\alpha_2};t\right)-G\left(-\tfrac{u}{v+\alpha_2};t\right)\Big]\,.
\end{split}\end{equation}
We see that there are no logarithmically divergent polylogarithms with argument $t$ and so we could just set $t=0$ to obtain the value of the integral at the upper bound. It is usually advisable to instead expand the polylogarithms to zeroth order around the desired point. This is more robust as it automatically enforces the cancellation of spurious logarithmic divergences or poles. This can be achieved using
\exbox{
    \inline{6}{& {\tt F2 = ExpandPolyLogs[Pp, \{t,0,0\}]; }}}
yielding the integrand for the last remaining integration
\begin{equation}
    \frac{\Delta_6\Big[G(-1;\alpha_2)-G(0;u)+G(0;v)-G(0;\alpha_2)+G(-v;\alpha_2)\Big]}{2\left(\alpha_2^2+(1-u+v)\alpha_2+v\right)}
\end{equation}
Here we see now a potential obstruction that is not automatically resolved by \plt{}: The denominator is quadratic in the next integration variable. In general this can be a fundamental obstruction, in the sense that that it may not be possible to evaluate the integral in terms of MPLs alone. In this case, however, there is only one integration left to do, so the result should be expressible as polylogarithms with algebraic arguments. \plt{} cannot automatically resolve this integral though, as it expects to be able to partial fraction the input into terms with only linear denominators. We can however manually solve the quadratic equation in the denominator
\begin{equation}
    \alpha_2 = \tfrac{1}{2}\left(u-v-1\pm\sqrt{(1-u-v)^2-4uv}\right)\,,
\end{equation}
and use it to reexpress the denominator in factorized form. At this point we are in principle done and could perform the next integral. We have to be careful to hide the square root from \texttt{Mathematica} as the routines for partial fractioning will otherwise restore the original quadratic form. Here we can just identify the root with the gram determinant $\Delta_6$ that we introduced before and thus define the new denominator as 
\exbox{
    \inline{7}{& {\tt De = 2*(1/2*(1-u+v-d6)+a[2]) *(1/2*(1-u+v+d6)+a[2]);}}}
 Now the next integration can be performed by invoking
 \exbox{\inline{8}{& {\tt P = GIntegrate[Numerator[F2]/De,a[2]];}}}
 The resulting primitive is a pure function of weight two that we refrain from spelling out here. Again we need to evaluate the primitive at the boundaries of integration. The lower boundary is immediately found using:
 \exbox{
     \inline{9}{& {\tt at0 = ExpandPolyLogs[P,{a[2],0,0}]}}\\
     \breakline\\
     \outline{9}{& {\tt 0}}}
In order to obtain the value at infinity, we once again need to invert the argument of the polylogarithms and use \texttt{ToFibrationBasis} to derive the required functional equations. Rather than calling \texttt{ToFibrationBasis} on the entire expression, we can also call it on a list of polylogarithms to obtain the individual functional equations. This can be particularly useful when we have a basis of functions for which we want to derive functional equations for that can then be stored. It is also useful if we want to inspect the individual functional equations to make sure that our choice of \texttt{FitValue} has not introduced any spurious $i\pi$ terms. In the present case there is the potential for such terms, as the polylogarithms contain spurious branchpoints for $u-v=0$ and as such the ordering of the variables that is employed when numerically fixing the branches of the logarithms becomes relevant. When we derive the functional equations in the following way we can easily verify that the obtain expressions maintain manifest reality:
\exbox{
    \inline{10}{& {\tt Pin = P/.a[2]->1/t;}}\\
    \inline{11}{& {\tt R = ToFibrationBasis[GetGs[Pin], \{t,d6,u,v\},}\\
    &{\tt\, FitValue->\{t->.1, d6->.12, u->.23, v->.45\}];}}}
%\exbox{
%    \inline{8}{& {\tt Pp = ToFibrationBasis[P/.a[2]->1/t,\{t,d6,u,v\}]}}}

We can then obtain the final result of the integral using
\exbox{
    \inline{12}{& {\tt Pin = Pin //. Dispatch[Thread[GetGs[Pin]->R]];}}\\
    \inline{13}{& {\tt result = ExpandPolyLogs[Pin,\{t,0,0\}]-at0;}}}

The four mass box is well known of course and its literature result is
\exbox{
    \inline{14}{& {\tt literature = PolyLog[2, ut] + PolyLog[2, vt] }\\
        & {\tt + 1/2*Log[u]*Log[v] - Log[ut]*Log[vt] - Zeta[2];}}}
with {\tt ut = 1/2*(1-u+v+d6)} and {\tt vt = 1/2(1+u-v+d6)}.

We can use the capability of \plt{} to compute numerical values for arbitrary polylogarithms to compare our result against the literature reference numerically:
\exbox{
    \inline{15}{& {\tt Ginsh[literature - result /.d6->Sqrt[(1-u-v)\^{}2-4*u*v],}\\
    & {\tt \{u->0.1,v->0.23\}]}}\\
    \breakline\\
    \outline{15}{& {\tt 0}}}

\section{Conclusion}
\label{sec:conclusion}
In this article we have reviewed the mathematical foundations of multiple polylogarithms and have documented their implementation in the \mathematica{} package \plt{}. Multiple polylogarithms have become ubiquitous in many areas of high-energy physics and the algorithms presented in this paper and implemented in \plt{} provide a powerful and flexible way to handle expressions involving multiple polylogarithms.
\plt{} has already served as the backbone of many recent calculations in high energy physics. Its public release accompanying this paper will enable even more studies of the mathematical structure of scattering amplitudes and calculations of multiloop amplitudes and cross sections. It also serves as a well-tested reference implementation for more specialized and optimized implementations that might be developed to handle particular situations.

\section*{Acknowledgments}
We would like to thank Herbert Gangl and Steven Charlton for collaboration on some of the mathematical aspects implemented into the code and for correspondence on the cobracket on MPLs. We also thank Lorenzo Tancredi 
and Brenda Penante for a careful reading of the manuscript. We would like to thank Enrico Herrmann and Andrew McLeod for suggesting the title and Enrico Hermann for testing the release version of the package.
This research was supported by the the ERC grant 637019 ``MathAm'', and the U.S.
Department of Energy (DOE) under contract DE-AC02-76SF00515.

%%%%%%%%%%%%%%%%%%%%%
% References 
%%%%%%%%%%%%%%%%%%%%%
%\bibliographystyle{elsarticle-num-names}
\bibliographystyle{JHEP}
\bibliography{polylogtools}

\end{document}